\documentclass{emulateapj}
\usepackage{color} 

\usepackage{epsfig}
\usepackage{subfigure}
\usepackage{graphicx}
\usepackage{mathrsfs}
\usepackage{amsmath}
\usepackage{natbib}
\newcommand{\pa}{\partial}
\newcommand{\mb}{\boldsymbol}

\shorttitle{Hall Effect in PPDs}
\shortauthors{X.-N. Bai}

\begin{document}


\title{Hall-effect Controlled Gas Dynamics in Protoplanetary Disks:
I. Wind Solutions at the Inner Disk}





\author{Xue-Ning Bai\altaffilmark{1}}
\affil{Institute for Theory and Computation, Harvard-Smithsonian
Center for Astrophysics, 60 Garden St., MS-51, Cambridge, MA 02138}
\email{xbai@cfa.harvard.edu}

\altaffiltext{1}{Hubble Fellow}





\begin{abstract}
The gas dynamics of protoplanetary disks (PPDs) is largely controlled by non-ideal
magnetohydrodynamic (MHD) effects including Ohmic resistivity, the Hall effect and
ambipolar diffusion. Among these the role of the Hall effect is the least explored and most
poorly understood. In this series, we have included, for the first time, all three non-ideal
MHD effects in a self-consistent manner to investigate the role of the Hall effect on PPD
gas dynamics using local shearing-box simulations. In this first paper, we focus on
the inner region of PPDs, where previous studies (Bai \& Stone 2013, Bai, 2013) excluding
the Hall effect have revealed that the inner disk up to $\sim10$ AU is
largely laminar, with accretion driven by a magnetocentrifugal wind. We confirm this basic
picture and show that the Hall effect introduces modest modifications to the wind
solutions, depending on the polarity of the large-scale poloidal magnetic field ${\mb B}_0$
threading the disk. When ${\mb B}_0\cdot{\mb\Omega}>0$, the horizontal magnetic field is
strongly amplified toward the disk interior, leading to a stronger disk wind (by $\sim50\%$
or less in terms of the wind-driven accretion rate). The enhanced horizontal field also leads
to much stronger large-scale Maxwell stress (magnetic braking) that contributes to a
considerable fraction of the wind-driven accretion rate. When ${\mb B}_0\cdot{\mb\Omega}<0$,
horizontal magnetic field is reduced, leading to a weaker disk wind (by $\lesssim20\%$)
and negligible magnetic braking. Moreover, we find that when ${\mb B}_0\cdot{\mb\Omega}>0$,
the laminar region extends farther to $\sim15$ AU before the magneto-rotational instability sets
in, while for ${\mb B}_0\cdot{\mb\Omega}<0$, the laminar region extends only to $\sim3$ AU
for a typical accretion rate of $\sim10^{-8}-10^{-7}M_{\bigodot}$ yr$^{-1}$. Scaling relations for
the wind properties, especially the wind-driven accretion rate, are provided for aligned and
anti-aligned field geometries. Issues with the symmetry of the wind solutions and grain
abundance are also discussed.
\end{abstract}



\section{Introduction}\label{sec:intro}

The gas dynamics in protoplanetary disks (PPDs) plays a crucial role in essentially
every aspect of planet formation. This is mainly because small dust grains are coupled
with the gas aerodynamically, while large solids are coupled with the gas gravitationally.
The global structure of the disk, as well as the level of turbulence are of particular 
importance. For example, the transport and growth of dust grains are sensitive to both
the radial pressure gradient and turbulence in PPDs (e.g.,
\citealp{Garaud07,Birnstiel_etal12,Pinilla_etal12,HughesArmitage12}),
the formation of planetesimals via collective effects such as streaming and gravitational
instabilities likely favors regions with small radial pressure gradient and low levels of
turbulence (e.g. \citealp{Johansen_etal09,BaiStone10b,BaiStone10c,Youdin11}),
dust grains may be trapped in vortices due to the Rossby wave instability at the pressure
maxima produced at inner dead zone edges
(e.g., \citealp{Lovelace_etal99,VarniereTagger06,Kretke_etal09}),
the growth of planetesimals into planetary cores may be suppressed when turbulence is
strong which will gravitationally excite their eccentricities, leading to destructive collisions
(e.g., \citealp{Ida_etal08,NelsonGressel10,Yang_etal12,OrmelOkuzumi13}),
and the migration of low to high mass planets, as well as gas accretion onto planetary
cores, are all sensitive to the radial disk structure as well as the local
microphysics (e.g., \citealp{Paardekooper_etal11,KretkeLin12,KleyNelson12,Gressel_etal13}).

The global structure of a PPD is mainly shaped by the process of angular momentum
transport, and the underlying mechanism largely dictates the level of turbulence in the disk.
Therefore, the key to understanding the gas dynamics of PPDs lies in determining the
mechanism of angular momentum transport, which is most likely magnetic in nature (see the
most recent review by \citealp{Turner_etal14}). The most important constraint on such
mechanisms comes from the fact that PPDs are actively accreting, with typical accretion rates
of $10^{-8\pm1}M_{\bigodot}$ yr$^{-1}$ \citep{Hartmann_etal98} over the lifetime of about 1-10
Myrs \citep{Sicilia_etal06,Ribas_etal14}, indicating efficient angular momentum transport must
take place in the entire disk.

Two leading mechanisms to transport angular momentum in accretion disks include the
magnetorotational instability (MRI, \citealp{BH91}) and the magnetocentrifugal wind
(MCW, \citealp{BlandfordPayne82}). The former generates strong turbulence, which
transports angular momentum radially outward within the disk as a viscous process (e.g.,
\citealp{ShakuraSunyaev73}), while the latter extracts angular momentum from the disk
vertically, which is then carried array by the wind. The details about whether and how
these mechanisms operate in PPDs largely depend on how gas and magnetic field
are coupled within the disk, as well as the geometry of the magnetic field.


Fully ionized gas can generally be well described by ideal magnetohydrodynamics (MHD)
where the gas and magnetic field are perfectly coupled with infinite conductivity. In contrast,
the extremely weakly ionized gas present in PPDs is subject to three non-ideal MHD effects:
Ohmic resistivity, the Hall effect, and ambipolar diffusion (AD). These
effects weaken the coupling between gas and magnetic fields in different ways, leading to
a reduced level of the MRI turbulence or even fully suppressing the MRI
\citep{Fleming_etal00,SanoStone02b,BaiStone11}. They also strongly affect the launching
process of the MCW \citep{WardleKoenigl93,Konigl_etal10,Salmeron_etal11}.

Calculations of ionization-recombination chemistry to infer the level of ionization
in PPDs demonstrate that all three non-ideal MHD effects are relevant and important in PPDs
\citep{Wardle07,Bai11a}. In particular, Ohmic resistivity
dominates in high densities with weak magnetic field,
applicable to the midplane region of the inner disk ($\lesssim10$AU). AD dominates in low
density regions with strong magnetic field,
applicable to the surface region of the inner disk, as well as the bulk of the outer disk
($\gtrsim30$AU). The Hall-dominated regime lies in between,
which covers a large fraction of PPDs, particularly the planet-forming regions.

The role of Ohmic resistivity has been the major focus for most works in the literature, and
has lead to the standard picture of layered accretion \citep{Gammie96}, followed by nearly
two decades of further developments from linear theory
\citep{Jin96,SanoMiyama99,Sano_etal00} to numerical simulations with increasing level of
complexity (e.g., \citealp{FlemingStone03,TurnerSano08,HiroseTurner11}). These works
have firmly established that the MRI does not operate in the midplane region of the inner disk
($\lesssim10$ AU) due to excessively large resistivity. This region is termed the dead zone.
Since Ohmic resistivity is completely negligible at the disk surface, the surface region is fully
MRI turbulent and is termed as the active layer.
The dead zone also has an inner edge ($<1$ AU) within which the MRI is activated due to
thermal ionization of Alkali species \citep{Fromang_etal02,Kretke_etal09,LatterBalbus12}.

Ambipolar diffusion (AD) is the second non-ideal MHD effect that receives considerable
attention \citep{BlaesBalbus94,KunzBalbus04,Desch04}. Non-linear simulations of the MRI
with AD (\citealp{BaiStone11}, in the ``strong-coupling" limit applicable to weakly ionized gas)
showed that in the AD-dominated regime, the MRI operates only when the magnetic field is
sufficiently weak with reduced level of turbulence. This finding led \citet{Bai11a,Bai11b} and
\citet{PerezBeckerChiang11a,PerezBeckerChiang11b} to conclude that MRI is insufficient to
drive rapid accretion at the observed rate of $10^{-8}M_{\bigodot}$ yr$^{-1}$ by at least
one order of magnitude, at least in the inner disk.

Recently, it has been demonstrated that when both Ohmic resistivity and AD are taken
into account with a self-consistent treatment of ionization-recombination chemistry, MRI is
either extremely inefficient or completely suppressed (depending on magnetic field geometry)
in the inner region of PPDs \citep{BaiStone13b,Bai13}. While this result seems surprising, it
is consistent with theoretical expectations, because the conventional ``active layer" is AD
dominated, and AD at the disk surface is strong enough to suppress the MRI. Without the MRI,
accretion is found to be efficiently driven by the MCW, and the desired rate of
$10^{-8}M_{\bigodot}$ yr$^{-1}$ can be easily achieved when the disk is threaded by some
weak net vertical magnetic field. Toward the outer disk where AD is expected to be the
sole dominant non-ideal MHD effect, MRI is able to operate; but to achieve sufficient accretion
rate, again net vertical magnetic flux is essential \citep{Simon_etal13a,Simon_etal13b}.
These results are pointing to a paradigm shift in our understanding of the gas dynamics in
PPDs, highlighting the importance of MCW and external magnetic field.

The Hall effect is the last non-ideal MHD effect yet to be included in self-consistent models
of PPDs. It has been shown to strongly affect the the linear properties of the MRI
\citep{Wardle99,BalbusTerquem01,WardleSalmeron12}. Non-linear simulations which
included both Ohmic resistivity and the Hall effect indicated that the Hall term changes the
saturation level of the MRI
\citep{SanoStone02a,SanoStone02b}. More recent simulations by \citet{KunzLesur13}
showed that when the Hall term is sufficiently strong, the system transitions into a ``low
transport state" characterized by a strong zonal field without transporting angular momentum.
These non-linear simulations highlight the potentially dramatic effect of the Hall term, and
raise concerns about the neglect of the Hall effect in most previous studies.

The Hall effect also affects the wind launching process hence the properties of the
magnetic wind, as studied in detail in \citet{Konigl_etal10} and \citet{Salmeron_etal11},
who extend the early work of \citet{WardleKoenigl93}. These authors identified the wind
launching criteria in the presence of all three non-ideal MHD effects separated in different
regimes and presented representative wind solutions. These works provided an important
theoretical framework for the general behavior of the wind solution. Their primary limitations
are unrealistic assumptions of constant Elsasser numbers (of order unity) and strong
vertical magnetic field (near equipartition at the midplane).

A special consequence of the Hall effect is that it makes the gas dynamics depend on
magnetic polarity: reversing the magnetic field would violate the original dynamical equations
and hence a different configuration is required. Since the Hall effect is prominent over a wide
range of disk radii, the gas dynamics of PPDs is largely Hall-controlled, and one expects it to
bifurcate into two branches with different field configurations and flow properties depending
on the polarity of the external magnetic field.

This paper, together with the companion paper, represent the first effort to explore the role
of the Hall effect in PPDs using non-linear MHD simulations with a self-consistent treatment
of the ionization-recombination chemistry. They serve as an extension of the recent work by
\citet{BaiStone13b} and \citet{Bai13} by further including the Hall effect.
In this first paper, we focus on the inner part of the disk ($R\lesssim10$ AU) where MRI is
expected to be suppressed over the entire vertical extent of the disk. We show that the
conclusion that the MRI is
suppressed with MCW-driven accretion still holds, while the property of the MCW is different
and depends on the polarity of the external large-scale magnetic field. We are aware of the
work of \citet{Lesur_etal14}, submitted the same time as the present paper, who emphasize
magnetic field amplification and enhanced magnetic braking due to the Hall effect. Our results
are consistent with theirs, while there are a number of differences which will be briefly
discussed. In the companion paper, we focus on the outer region of PPDs and address
how the behavior of the MRI is affected by the Hall effect.

This paper is organized as follows. Given the increasing level of complexity compared with
previous works, especially involving the full non-ideal MHD physics, we devote Section 2 to
background information intended to guide the readers through the formulation and the role
played by individual non-ideal MHD effects, highlighting the new features introduced by the
Hall term. Section 3 describes the methodology of our numerical simulations as well as the
simulation runs. In Section 4, we focus on a particular set of simulations with fiducial parameters
and discuss how the Hall effect modifies the original wind solution obtained by \citet{BaiStone13b}
and the properties of the new wind solutions. We then extend the results with a much broader
range of parameters in Section 5. In Section 6 we discuss the implications of our findings
and conclude.

\section[]{Preliminaries}

\subsection[]{Disk Model}

We plan to study the local gas dynamics of PPDs across a wide range of disk radii. Since we
are interested in short timescales ($\sim100$ local orbital time, compared with the disk lifetime),
we adopt a fixed disk model without worrying about global disk evolution. As a convention,
we use the minimum-mass solar nebular (MMSN) disk as our standard model, with
surface density and temperature given by \citep{Weidenschilling77,Hayashi81}
\begin{equation}
\begin{split}
\Sigma(R)&=1700R_{\rm AU}^{-3/2}\ {\rm g\ cm}^{-2}\ ,\\
T(R)&=280R_{\rm AU}^{-1/2}\ {\rm K}\ .
\end{split}
\end{equation}
where $R_{\rm AU}$ is disk radius measured in AU. We treat the disk as vertically isothermal,
with isothermal sound speed given by $c_s=0.99R_{\rm AU}^{-1/4}\ {\rm km\ s}^{-1}$ (mean
molecular weight $\mu=2.34m_p$). While in reality the disk is hotter at the surface and colder
in the midplane due to stellar irradiation, we are mainly interested in the role played by magnetic
fields which is likely the primary driving force of disk angular momentum transport, and we leave
more realistic treatment of thermodynamics for future work.

\subsection[]{Formulation}\label{ssec:formula}

We study the gas dynamics in PPDs using the standard local shearing-sheet approximation
\citep{GoldreichLyndenBell65}, where MHD equations are written in Cartesian coordinates in
the corotating frame at a fiducial radius $R$ with Keplerian frequency $\Omega$. The radial,
azimuthal, and vertical dimensions are represented by $x, y$ and $z$ coordinates.
Background Keplerian shear ${\mb u}_0=-(3/2)\Omega x{\mb e}_y$ is subtracted from the
formulation, with $\rho$ and ${\mb v}$ denoting gas density and (background shear subtracted)
velocity, respectively. Including the stellar vertical gravity, the equations read
\begin{equation}
\frac{\pa\rho}{\pa t}+\nabla\cdot(\rho{\mb v})+u_0\frac{\pa\rho}{\pa y}=0\ ,\label{eq:cont}
\end{equation}
\begin{equation}
\begin{split}
&\frac{\pa{\mb v}}{\pa t}+({\mb v}\cdot\nabla){\mb v}+u_0\frac{\pa{\mb v}}{\pa y}=\\
&-\frac{\nabla P}{\rho}+\frac{{\mb J}\times{\mb B}}{\rho}-\frac{1}{2}\Omega v_x{\mb e}_y
+2\Omega v_y{\mb e}_x-\Omega^2z{\mb e}_z\ ,\label{eq:momentum}
\end{split}
\end{equation}
where ${\mb B}$ is the magnetic field, whose unit is such that magnetic permeability is
1, and ${\mb J}=\nabla\times{\mb B}$ is the current density.
We use an isothermal equation of state with $P=\rho c_s^2$. In hydrostatic equilibrium, the
gas density profile follows $\rho=\rho_0\exp{(-z^2/2H^2)}$, where $\rho_0$ is the midplane
gas density, and $H\equiv c_s/\Omega$ is the disk scale height.

For very weakly ionized gas as in PPDs, the above single-fluid equations describe the
dynamics for the bulk of the neutral gas. Note that the neutral gas also feels the Lorentz
force, which is effectively achieved by colliding with charged particles.

The charged particles contain negligible inertia, and in the dense environment of PPDs
(collision frequency with the neutrals is much higher than orbital frequency), their
dynamics is fully determined by the balance between Lorentz force and collisional
drag with the neutrals. In this so-called ``strong coupling" limit, multi-fluid equations are
unnecessary. The motion of charged particles simply provides the conductivity for the
bulk of the gas, which is generally anisotropic due to the presence of magnetic field.
Reflecting to the induction equation, such anisotropic conductivity introduces three
non-ideal MHD effects in addition to the normal inductive term

\begin{equation}
\frac{\pa{\mb B}}{\pa t}
=\nabla\times({\mb v}\times{\mb B})-\frac{3}{2}B_x\Omega{\mb e}_y
-\nabla\times{\mb E}'\ ,\label{eq:induction}
\end{equation}
with
\begin{equation}
{\mb E}'\equiv\eta_O{\mb J}+\eta_H({\mb J}\times{\hat{\mb B}})
+\eta_A{\mb J}_\perp\ ,\label{eq:niemf}
\end{equation}
where ${\mb E}'$ is the electric field (in the comoving frame) due to non-ideal MHD
terms, $\hat{\mb B}$ denotes the unit vector along ${\mb B}$, subscript ``$_\perp$"
denotes the vector component perpendicular to ${\mb B}$, and $\eta_O, \eta_H$
and $\eta_A$ are the Ohmic, Hall and ambipolar diffusivities. The total electric field is
\begin{equation}
{\mb E}=-{\mb v}\times{\mb B}+{\mb E}'\ .\label{eq:emf}
\end{equation}

The general expression of these diffusivities involve the abundance of all charged
species \citep{Wardle07,Bai11a},
but in the absence of small charged grains, the diffusivities can be cast into a particularly
simple form\footnote{Equation (\ref{eq:diff0}) is written in Gaussian units, for ease of
comparison to standard expressions.}:
\begin{equation}
\begin{split}
&\eta_O=\frac{c^2m_e\gamma_e\rho}{4\pi e^2n_e}\propto\bigg(\frac{n_H}{n_e}\bigg)\ ,\\
&\eta_H=\frac{cB}{4\pi en_e}\ \ \propto\bigg(\frac{n_H}{n_e}\bigg)\bigg(\frac{B}{\rho}\bigg)\ ,\\
&\eta_A=\frac{B^2}{4\pi\gamma_i\rho\rho_i}\propto\bigg(\frac{n_H}{n_e}\bigg)\bigg(\frac{B}{\rho}\bigg)^2\ ,
\end{split}\label{eq:diff0}
\end{equation}
where $n_H$ is the number density of hydrogen nuclei, $\gamma_e$ and $\gamma_i$
denote coefficients of momentum transfer for electron-neutral and ion-neutral collisions
(see \citealp{Bai11a}), $n_e$ is the electron number density and $\rho_i$ is the ion mass
density. We define $n_e/n_H$ as the ionization fraction.
In this largely grain-free case, the Ohmic resistivity describes collisions between electrons
and neutrals, the Hall term describes the electron-ion drift ${\mb v}_e-{\mb v}_i$, and the
AD term describes the ion-neutral drift ${\mb v}_i-{\mb v}$.
We further see that the strength of all three effects is inversely proportional to the ionization
fraction $n_e/n_H$, while their dependence on $(B/\rho)$ reveals that Ohmic resistivity
(independent of $B/\rho$) dominates in dense regions with weak magnetic field, AD
dominates in sparse regions with strong magnetic field, and the Hall-dominated regime lies
in between.

The importance of these non-ideal MHD effects in PPDs can be characterized by defining
an Elsasser number for each term
\begin{equation}\label{eq:Elsasser}
\Lambda\equiv\frac{v_A^2}{\eta_O\Omega}\ ,\quad
\chi\equiv\frac{v_A^2}{\eta_H\Omega}\ ,\quad
Am\equiv\frac{v_A^2}{\eta_A\Omega}\approx\frac{\gamma_i\rho}{\Omega}\ ,
\end{equation}
where $v_A=\sqrt{B^2/\rho}$ is the Alfv\'en velocity. The non-ideal MHD terms become
dynamically important when any of these Elsasser numbers become much smaller than $1$,
while the ideal MHD limit applies when they largely exceed 1. Note that $Am$ is independent
of magnetic field strength, and in the absence of small grains, it corresponds to the number of
times a neutral molecule collides with the ions in a dynamical time ($\Omega^{-1}$).

\subsection[]{Hall Effect and Characteristics}

Working with Equation (\ref{eq:diff0}) for magnetic diffusivities, we can first define the
Hall frequency as
\begin{equation}
\omega_H\equiv\frac{en_eB}{m\rho c}=\frac{\rho_i}{\rho}\omega_i\ ,\label{eq:omgH}
\end{equation}
where $\omega_i=eB/m_ic$ is the gyro-frequency of the ions. Therefore, the Hall
frequency is simply the ion gyro-frequency reduced by the level of ionization. With
this definition, the Hall Elsasser number is simply given by
\begin{equation}
\chi=\frac{\omega_H}{\Omega}\ .
\end{equation}

The Hall effect is not dissipative because the Hall electric field
${\mb E}'_H\propto{\mb J}\times{\mb B}$ is perpendicular to ${\mb J}$. Instead
of dissipation, the Hall effect breaks the degeneracy between left and right
polarized Alfv\'en waves. The left-handed wave does not propagate beyond
$\omega_H$, while the right-handed wave (the whistler wave) has asymptotic
dispersion relation $\omega\propto k^2$ at
$\omega\gg\omega_H$ (see Appendix B and Equation (\ref{eq:whistlerDR2})).
We see that the Hall effect is important on timescales comparable to or shorter than
$\omega_H^{-1}$, where the whistler wave physics comes into play. Since the
gas dynamics in PPDs is characterized by dynamical timescale
$\Omega^{-1}$, the Elsasser number characterizes the importance of the
Hall term well.

Unlike Ohmic resistivity and AD, the effect of the Hall term depends on magnetic
polarity. In the induction equation (\ref{eq:induction}), if one reverses the magnetic
field, the Hall term does not change sign while all other terms do. Hence, the Hall
term breaks the magnetic reversal symmetry which holds broadly in ideal/resistive/AD
MHD\footnote{In the shearing-sheet approximation, a steady-state wind solution is
always invariant under the transformation ${\mb B}_h\rightarrow-{\mb B}_h$,
${\mb v}'_h\rightarrow-{\mb v}'_h$, where subscript `$_h$' denotes the horizontal
component. Without the Hall term, the wind solution is also invariant under
$B_z\rightarrow-B_z$, ${\mb v}'_h\rightarrow-{\mb v}'_h$.}. In PPDs, this means
that the gas dynamics is expected to be different when the external magnetic field
is aligned or anti-aligned with ${\mb\Omega}$. For our choice, the aligned and
anti-aligned cases correspond to background net vertical field $B_{z0}>0$ and
$B_{z0}<0$ in shearing-box simulations.

\subsection[]{MRI Suppression and Disk Wind Launching}\label{ssec:mriwind}

The focus of this paper is the inner region of PPDs, where we expect the MRI to
be suppressed and a disk wind to launch. The two facts are closely related, and
depend on the amount of external vertical magnetic flux threading the disk.
This net vertical field $B_{z0}$ is characterized by the parameter $\beta_0$
\begin{equation}\label{eq:beta0}
\beta_0\equiv\frac{P_{g,{\rm mid}}}{P_{B0}}=\frac{\rho_0c_s^2}{B_{z0}^2/2}\ .
\end{equation}
Here we use subscript `$_0$' to specifically denote the background values. The
plasma $\beta$ defined using total field strength can be much smaller.

In the ideal MHD limit, the MRI operates efficiently for $\beta_0\gtrsim100$,
where stronger net field gives stronger turbulence \citep{BaiStone13a}.
Further increasing the net vertical flux would stabilize the MRI, which is not expected
to operate for $\beta_0\lesssim10$ (e.g., \citealp{Latter_etal10,Lesur_etal13}).
Non-ideal MHD effects modify the properties of the MRI in different ways, as summarized
in Section \ref{sec:intro}.
In the inner region of PPDs ($\lesssim10$ AU), it was found that the threshold for MRI
suppression switches to much weaker field $\beta_0\sim10^{5-6}$ \citep{BaiStone13b,Bai13}.
This is because of the excessively large resistivity around the disk midplane, and strong
AD at the disk surface.

From local shearing-box simulations of the MRI, it was found that the presence of
net vertical magnetic field always leads to launching of a disk outflow
(e.g., \citealp{SuzukiInutsuka09,OkuzumiHirose11,Fromang_etal13,BaiStone13a}).
The outflow is magnetocentrifugal in nature \citep{BlandfordPayne82}, but it is
unclear whether it connects to a global magnetocentrifugal wind mainly because
of the MRI dynamo activities and symmetry issues \citep{BaiStone13a}. Most recent
global MRI simulations are still inconclusive on the fate of such a disk outflow due to
limited vertical domain size and other numerical issues \citep{SuzukiInutsuka14}.

Launching of a steady disk wind generally requires the presence of strong net
vertical field with $\beta_0\sim1$ (e.g., \citealp{WardleKoenigl93,FerreiraPelletier95}),
which is also found to be the case from local steady state wind solutions that include
all non-ideal MHD effects \citep{Konigl_etal10,Salmeron_etal11}. However, these
conditions are all derived by assuming constant magnetic diffusivities or constant
Elsasser numbers and the wind is essentially launched from the disk interior.
More appropriately, launching of a disk wind only requires equipartition field at the wind
launching region (e.g., \citealp{Li96,Wardle97}). In the inner region of PPDs, the
disk interior is essentially decoupled from the magnetic field due to excessively large
Ohmic resistivity, therefore, wind launching is only possible from the disk surface
layer where gas and magnetic fields are better coupled. Since equipartition field at the
low density disk surface corresponds to much larger $\beta_0$, a steady wind can be
naturally launched with $\beta_0\sim10^5$ \citep{BaiStone13b,Bai13}.

In brief, strong non-ideal MHD effects of Ohmic resistivity and AD in the inner region
of PPDs makes the launching of steady disk winds much easier, and can be achieved
with very weak net vertical field. This is closely related to the suppression of the MRI
discussed earlier since MRI is the main source that prevents launching a steady wind.

\subsection[]{Structure and Symmetry of the Wind Solution}\label{ssec:symmetry}

The wind solutions presented in this paper extend earlier wind solutions of
\citet{BaiStone13b} and \citet{Bai13} by including the Hall term, and they share many
common properties. The wind launching process is described in Figure 6 and
Section 4.1 of \citet{BaiStone13b}.
For the laminar wind solution, we can divide the disk vertical extent into a disk zone
containing the disk midplane where the azimuthal gas velocity is sub-Keplerian, and a
wind zone at the disk surface where the azimuthal velocity is super-Keplerian. The height
at which this transition occurs, $z_b$, is referred to as the base of the wind
\citep{WardleKoenigl93}.

The wind carries away disk angular momentum, the rate of which is determined
by the $z\phi$ component of the stress tensor $T_{z\phi}$ at the base of the wind $z_b$
\citep{BaiStone13b}
\begin{equation}
T_{z\phi}^{z_b}=-B_zB_y|_{\pm z_b}
\end{equation}
Note that only the Maxwell (magnetic) component is involved, because the Reynolds
(hydrodynamic) component is simply zero at $z_b$ by definition. The value of
$z_b$ is typically found to be $\sim4H$ or higher in the inner disk.

The total rate of angular momentum loss from the disk is given by the difference of the
above stress at the top and bottom of the disk, $\pm z_b$. Since $B_z=B_{z0}$ is
constant throughout the disk, the desired symmetry for the wind to extract disk
angular momentum is the even-$z$ symmetry, where
\begin{equation}
B_{x,y}(z)=-B_{x,y}(-z)\ ,\quad v_{x,y}(z)=-v_{x,y}(-z)\ ,
\end{equation}
hence the radial field bends to the same direction at the top and
bottom of the disk. Correspondingly, the rate of wind-driven accretion is given by
\begin{equation}\label{eq:dotMV}
\dot{M}_{\rm V}=\frac{8\pi}{\Omega}R|T_{z\phi}^{z_b}|\approx
4.1\times10^{-8}M_{\bigodot}\ {\rm yr}^{-1}
\bigg(\frac{|T_{z\phi}^{z_b}|}{10^{-4}\rho_0c_s^2}\bigg)R_{\rm AU}^{-3/4}\ ,
\end{equation}
where subscript '$_{\rm V}$' represents accretion driven by vertical angular
momentum transport, and in the latter estimate, we have adopted the MMSN
disk model, with $R_{\rm AU}$ being disk radius normalized to AU.

In numerical simulations containing both sides of the disk, it was found that the
simulations sometimes generate solutions with odd-$z$ symmetry
($B_{x,y}(z)=B_{x,y}(-z)$, $v_{x,y}(z)=v_{x,y}(-z)$), which is unphysical for a disk wind
since it implies that the radial field at the top and bottom bending to opposite directions.
This may be due to limitations of the shearing-box framework, where disk curvature is
ignored, and hence there is no distinction between inward or outward radial directions.
Detailed discussions can be found in Section 4.4 of \citet{BaiStone13b}, where it was
found that with Ohmic resistivity and AD included, a physical solution can be obtained
and maintained by flipping the horizontal field at one side of the disk. However, the
physical solution does not strictly obey the even-$z$ symmetry: the flip does not
exactly take place at the disk midplane, but at some height above through a thin
layer. This is because the midplane region is too resistive to conduct electric
current, and only in the upper layer (typically at $z\sim1-3H$) can the flip take place
where there is marginal coupling between gas and magnetic field.
The thin layer carries a strong current, and receives the entire Maxwell stress from
the wind. Correspondingly, it possesses large radial velocities and carries the entire
accretion flow.

Despite this issue with the symmetry of the wind solution, it was found that the solution
in the wind zone, in particular, $T_{z\phi}^{z_b}$, is independent of such symmetry
(see Section 4.4.1 of \citealp{BaiStone13b} for details). This is mainly because $z_b$ is
typically higher than the location where horizontal field flips. For this reason, if we
are mainly interested in the properties of the disk wind, it suffices to enforce the
even-$z$ symmetry by simulating half the disk ($z\geq0$) with reflection boundary
condition.

\subsection[]{Radial Transport of Angular Momentum}\label{ssec:dLinR}

Besides vertical extraction of angular momentum via disk wind, angular momentum
can be transported radially outward within the disk, characterized by the $R\phi$
component of the stress tensor $T_{R\phi}$
\begin{equation}
T_{R\phi}\equiv T_{R\phi}^{\rm Rey}+T_{R\phi}^{\rm Max}
=\overline{\rho v_xv'_y}-\overline{B_xB_y}\ ,
\end{equation}
where the overline represents horizontal average. The Shakura-Sunyaev $\alpha$
is obtained by vertically integrating $T_{R\phi}$ across the disk zone
\begin{equation}
\alpha\equiv\frac{\int_{-z_b}^{z_b}T_{R\phi}dz}{c_s^2\int\rho dz}\ .\label{eq:alpha}
\end{equation}

Assuming steady state accretion, the accretion rate resulting from radial
angular momentum transport then reads
\begin{equation}
\dot{M}_{\rm R}=\frac{2\pi}{\Omega}\alpha c_s^2\Sigma
\approx0.82\times10^{-8}M_{\bigodot}{\rm yr}^{-1}\bigg(\frac{\alpha}{10^{-3}}\bigg)R_{\rm AU}^{-1/2}\ ,\label{eq:dotMR}
\end{equation}
where subscript `$_{\rm R}$' represents accretion driven by radial angular
momentum transport, and in the second equation we have adopted the MMSN
disk model.

In the case of MRI turbulence, $T_{R\phi}^{\rm Max}$ and $T_{R\phi}^{\rm Rey}$
are typically dominated by contributions from turbulence, while in MRI inactive
regions, substantial Maxwell stress can be achieved due to large-scale magnetic
field $-\overline{B}_x\overline{B}_y$ \citep{TurnerSano08}. Such large-scale field
corresponds to ordered horizontal magnetic field that winds up into spirals and
transports angular momentum outward by means of {\it magnetic braking}.
Since we consider pure laminar wind solutions, radial transport of angular momentum
is almost completely due to magnetic braking (\citealp{BaiStone13b}, and Reynolds
stress is typically negligible).

\section[]{Simulation Setup and Parameters}

\subsection[]{Method}

We use ATHENA, a higher-order Godunov MHD code with constrained
transport technique to enforce the divergence-free constraint on the
magnetic field \citep{GardinerStone05,GardinerStone08,Stone_etal08}
for all calculations presented in this paper. Non-ideal MHD terms
including Ohmic resistivity \citep{Davis_etal10}, and AD \citep{BaiStone11}
have been developed for Athena. In this work, we have further implemented
the Hall term, with detailed algorithms described in Appendix A, and
code tests shown in Appendix B. Following the formulation in Section
\ref{ssec:formula}, all our simulations are carried out using the shearing-box
module with orbital advection \citep{StoneGardiner10}. We use the HLLD
Riemann solver \citep{MiyoshiKusano05} with third order reconstruction.
Outflow vertical boundary condition is used, where gas density is
extrapolated assuming hydrostatic equilibrium, zero-gradient is assumed for
velocity and magnetic field except that $v_z$ in the ghost zones is set to zero
if the flow is ingoing at the boundary. We also replenish disk mass to
compensate for mass loss, although the mass loss is negligible over the
duration of most simulations. We always adopt natural unit in the simulations
with $\rho_0=c_s=\Omega=1$.

All our simulations are quasi-1D along the disk vertical dimension to construct
laminar wind solutions. They are quasi-1D because we use a three-dimensional
(3D) simulation box with only $4\time4$ cells in the horizontal dimensions. The
additional horizontal dimensions were found to be necessary for our time-dependent
simulation to properly relax to the laminar wind configuration \citep{BaiStone13b}.
For most of our runs, the vertical domain covers half of the disk, extending from
$z=0$ to $z=8H$. Reflection boundary condition at $z=0$ is enforced to achieve
the desired even-$z$ symmetry for physical wind solutions. We also perform a
few simulations with full disk from $z=-8H$ to $8H$ to address issues with
symmetry and the strong current layer. In the vertical dimension, we use a
resolution of 24 cells per $H$, which we find to be sufficient to properly resolve
the wind structure (reducing this resolution by a factor of 2 yields essentially the
same wind solution).

The magnetic diffusivities $\eta_O$, $\eta_H$ and $\eta_A$ are obtained
self-consistently in the simulations based on a pre-computed look-up table
assuming equilibrium chemistry. For $\eta_H$ and $\eta_A$, they are given
in $\eta_H/B$ and $\eta_A/B^2$ which are independent of $B$ for the
regimes we consider in this paper (in the absence of abundant small grains).
Since we adopt the MMSN disk model with
isothermal equation of state, the diffusivity table is two-dimensional providing
the diffusivities as a function of density and ionization rate at fixed temperature.
The ionization rate includes contributions from stellar X-ray, cosmic rays and
radioactive decay, are expressed a function of column density to the disk surface
(see Section 3.2 of \citealp{Bai11a}), where fiducially we adopt X-ray luminosity
of $L_X=10^{30}$ erg s$^{-1}$ and X-ray temperature of 5 keV\footnote{The X-ray
ionization rate calculations have been updated recently by \citet{ErcolanoGlassgold13}
who found results consistent with previous calculations of \citet{IG99}
which we adopt.}. The procedure closely follows the description in
\citet{BaiStone13b}, with some changes and updates described below.

We have updated our chemical reaction network with the most recent version
of the UMIST database \citep{UMIST12}. Reactions are extracted using the
same list of chemical species adopted in our previous works
\citep{BaiGoodman09,Bai11a,Bai11b,BaiStone13b,Bai13}, which originated
from the work of \citet{IlgnerNelson06}. The total number of gas-phase reactions
increases from 2083 to 2147. The grain-binding energy of all species are also
updated to new values. We have tested the new chemical network and found
that for a grain-free calculation, the new network gives ionization fractions
that are typically slightly smaller compared with the previous version, but within
a factor of $2$. Fiducially, we include a single population of dust grains with
size $a=0.1\mu$m and abundance of $10^{-4}$ in mass, which is the same as
used in our earlier works \citep{BaiStone13b,Bai13}. While this is by no means
realistic, it provides reasonable and representative amount of total surface area
to enhance recombination. It has been shown that the properties of wind
solutions depend very weakly on the grain abundance \citep{BaiStone13b}, mainly
because the wind is launched from disk upper layers where ionization fraction
$\gg$ grain abundance.

\begin{figure*}
    \centering
    \includegraphics[width=180mm]{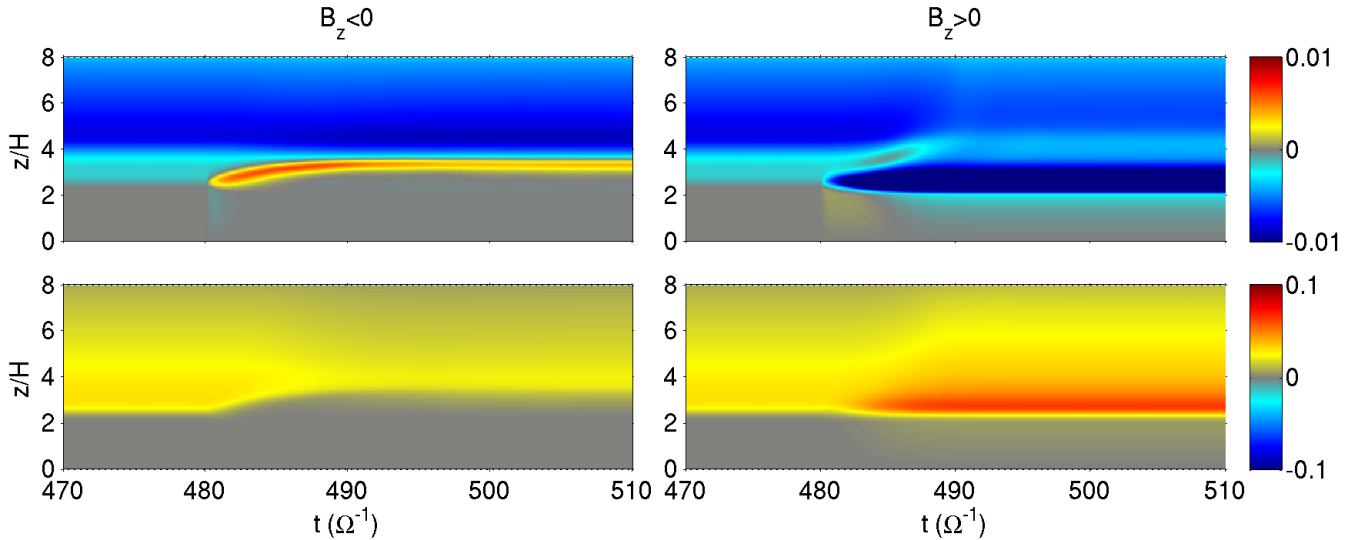}
  \caption{Time evolution of the magnetic field profile $B_x$ (upper panels) and $B_y$
  (lower panels) around the time the Hall effect is turned on at $t=480\Omega^{-1}$.
  Left and right panels correspond to cases with $B_z<0$ and $B_z>0$ respectively,
  where we have properly flipped the magnetic field in the case of $B_z<0$ to enable
  more direct comparison.
  }\label{fig:relax}
\end{figure*}

The disk surface layer is also exposed to far-UV (FUV) radiation which greatly
enhances the level of ionization (but is not captured in our diffusivity table) so
that the gas behaves in the ideal MHD regime. In our previous works
\citep{BaiStone13a,Bai13}, we obtained the diffusivities in the FUV layer separately
by assuming constant ionization fraction of $\sim10^{-5}-10^{-4}$, and the FUV
layer was assumed to have a penetration depth of $0.01-0.1$ g cm$^{-2}$ based on
the work by \citet{PerezBeckerChiang11b}. Correspondingly, there is a sharp jump
of diffusivities across the FUV ionization front (see the lower left panel of Figure 5 in
\citealp{BaiStone13b}). More self-consistent X-ray and UV radiative-transfer
calculations (e.g., \citealp{Walsh_etal10,Walsh_etal12}) showed that the ionization
fraction increases smoothly from midplane to surface. To avoid unrealistically
sharp transitions,
we empirically treat the FUV ionization as another independent ionization source,
with ionization rate of
\begin{equation}
\xi_{\rm FUV}=1.0\times10^{-6}R_{\rm AU}^{-2}\exp{(-\Sigma/\Sigma_{\rm FUV})}
{\rm s}^{-1}\ .
\end{equation}
The ionization is assumed to act on hydrogen and helium in the same way as X-ray
and cosmic-rays so that we can simply use the diffusivity table by extending it to
higher ionization rates. This assumption is by no means physical (FUV ionization
does not act on H or He), but it works for our purpose, because we simply need a
prescription to allow the gas to behave in the ideal MHD regime in the FUV layer
with a smooth transition. The detailed ionization structure in the FUV layer is
unimportant. To further validate this choice, we have calculated the ionization
profiles at 1, 10 and 100 AU based on the above ionization rate at the disk surface,
and compared the results with the radiative transfer and chemistry calculations of
\citet{Walsh_etal12} with the same X-ray luminosity and temperature\footnote{We
sincerely acknowledge H. Nomura and C. Walsh for rerunning their calculations
with new parameters and providing us the data for comparison.}. We find reasonable
agreement when $\Sigma_{\rm FUV}\approx0.005$ g cm$^{-2}$, which will be the
standard value we adopt in this paper.

From our chemistry calculations, the magnetic diffusivities at the disk midplane can
become excessively large, leading to extremely small timesteps from the Courant
condition. Besides using super time-stepping to handle Ohmic resistivity and AD
(see Appendix A), we further set a diffusivity floor $\eta_{\rm flr}=10c_sH$ so that
$\eta_O+\eta_H+\eta_A\leq\eta_{\rm flr}$. If the floor value is reached, the values of
$\eta_O$, $\eta_H$ and $\eta_A$ are reduced proportionally so as not to affect their
relative importance. We have verified that this floor value is sufficiently large and the
properties of our wind solutions are independent of $\eta_{\rm flr}$\footnote{Increasing
the floor value by a factor of 3 has no influence to our fiducial run R1b5H+, while
for our run with full box R1b5H+Full, the value of $\alpha^{\rm Max}$ is increased by
$\lesssim15\%$.}.

\begin{figure*}
    \centering
    \includegraphics[width=180mm]{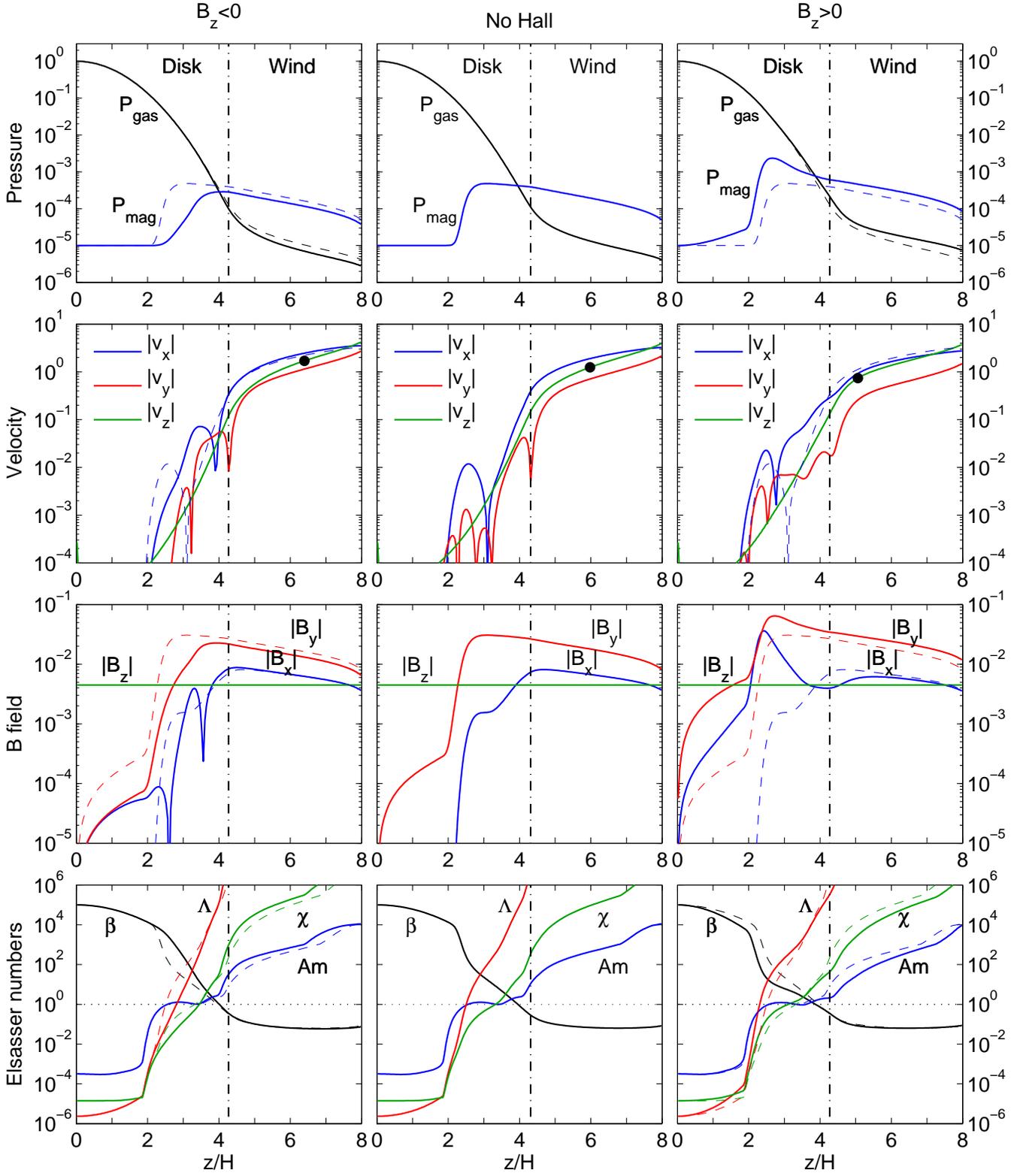}
  \caption{Vertical profiles of various quantities in the laminar wind solutions from
  our fiducial simulations (R=1AU, $\beta_0=10^5$), where reflection symmetry about
  the midplane is enforced (for physical wind geometry). Left panels: solution with
  $B_z<0$ (solid lines). Middle panels: solution without including the Hall effect. Right
  panels: solution with $B_{z}>0$ (solid line). Top row: gas and magnetic pressure; 
  second row: three velocity components, where the black dot marks the Alfv\'en point;
  third row: three magnetic field components; bottom row: Ohmic, Hall and AD Elsasser
  numbers, as well as the total plasma $\beta$. For comparison, we also show the
  Hall-free solution in dashed lines in the left and right panels (but only $|v_x|$ in the
  second row to avoid confusion). In all panels, the vertical dash-dotted
  line marks the location of the base of the wind $z=z_b$. Velocity and magnetic field
  components are shown for absolute values. For velocities, all components are positive
  in the wind zone, the sign changes every time the curve undergoes a kink in
  the logarithmic plot. For magnetic fields, $B_x, -B_y$ and $B_z$ have the same sign
  in the wind zone. Note the sign of $B_x$ flips twice in the $B_z<0$ simulation.
  }\label{fig:fiducial}
\end{figure*}

\subsection[]{Simulation Runs}

Our simulations mainly have two parameters, namely, the radial location in the disk
$R$, and the net vertical magnetic field $B_z$ characterized by $\beta_0$. For each
combination of the two parameters, we have three simulation runs. We begin by
including only the Ohmic resistivity and AD,
running to $t=480\Omega^{-1}$ where the system has fully settled into a laminar
configuration. We then turn on the Hall term and split the simulation into two more runs:
one is continued from the first run, with $B_z>0$ (aligned), and for the other we flip
all three components of the magnetic field (anti-aligned), while keeping the velocities
unchanged. These simulations are run for another $\sim80$ orbits to
$t=960\Omega^{-1}$, which is sufficient for the system to relax to a new wind solution.

Our simulation runs are named as run R$x$b$y$H$*$, where $x$ represents disk
radius in AU, $y=\log_{\rm 10}\beta_0$, and $*$ can be $0$, `$+$', `$-$' denoting initial
simulation without the Hall term ($0$), continued simulation with Hall term for $B_z>0$
(`$+$') and $B_z<0$ (`$-$'), respectively. For instance, in Section \ref{sec:fid}, we
focus on our fiducial runs R1b5H$*$, which are fixed at
$R=1$AU with $\beta_0=10^5$. We show in Figure \ref{fig:relax} the time evolution of
horizontal magnetic field profiles around the time the Hall effect is turned on. The steady
state profile prior to $t=480\Omega^{-1}$ belongs to run R1b5H0, after which the two
runs evolve differently due to different field polarities. In Section \ref{ssec:issue}, we also
perform simulations with full vertical domain to address the symmetry issues, named as
R1b5H$*$Full. These runs and results and listed in Table \ref{tab:runs1}.

In Section \ref{sec:param},  we first consider the fiducial runs with variations in other
parameters, also listed in Table \ref{tab:runs1}. The variations are labeled by attaching
additional letters in front of the standard run names. We consider disk masses that are
$3$ and $0.3$ times the MMSN disk, labeled by `M3' and `M03'. We also perform runs
with grain-free chemistry, labeled by `nogr'. Finally, we vary the X-ray ionization rate to
$L_X=10^{29}$ and $10^{31}$ ergs s$^{-1}$, labeled by `X29' and `X31'.
In the remainings of Section 5, we further consider runs with $\beta_0=10^4$ and
$10^6$, and $R$ at 0.3, 3, 5, 8 and 15 AU, where all other parameters are fixed at
standard values. The list of these simulation runs are provided in Table \ref{tab:runs2}.

\section[]{Simulation Results: Representative Wind Solutions}\label{sec:fid}

We begin by focusing on a fiducial set of simulations at fixed radius of 1 AU with
$\beta_0=10^5$.
In Figure \ref{fig:fiducial}, from left to right, we show the general properties of the wind
solutions for runs R1b5H--, R1b5H0 and R1b5H+ respectively. Major diagnostic
quantities of these solutions are provided in Table \ref{tab:runs1}. The rest of this
section is devoted to discussing the properties of these solutions.

\subsection[]{Relaxation to New Wind Solutions}\label{ssec:relax}

We start from the middle panels of Figure \ref{fig:fiducial} for run R1b5H0, where
the Hall term was not included. The solution closely resembles the fiducial
solution in our previous work \citet{BaiStone13b} (see their Figures 5, 11 for
solutions with odd and even symmetries), except that the Elsasser numbers
in this work increase smoothly to the surface FUV layer as a result of the
new procedure adopted in this work. While the depth of the FUV layer
is uncertain, such smooth transition is likely more realistic.

Solutions including the Hall effect (R1b5H$\pm$) are shown in the left and right
panels of Figure \ref{fig:fiducial}. We see that the main effect is that the horizontal
magnetic field is strongly amplified when $B_z>0$, while the field is largely reduced
when $B_z<0$. The amplification and reduction mainly result from the
Hall-dominated region at $z\sim2-3H$. Moreover, for $B_z<0$, the sign of $B_x$
reverses in the Hall-dominated region so that it has the same sign as $B_y$. In
our time-dependent simulations, relaxation from the original solution R1b5H0 to the
new solutions R1b5H$\pm$ is very rapid, as we see in Figure \ref{fig:relax}. The
initial evolution of $B_x$ takes less than an orbit, with a few more orbits
to fully relax to the final configuration.

These features can be understood by looking at the induction equation. With the
addition of the Hall term, the immediate evolution of magnetic field follows
\begin{equation}
\frac{\pa B_x}{\pa t}=\eta_{Hz}\frac{\pa^2}{\pa z^2}B_y\ ,\label{eq:hallshr1}
\end{equation}
\begin{equation}\label{eq:hallshr2}
\frac{\pa B_y}{\pa t}=-\eta_{Hz}\frac{\pa^2}{\pa z^2}B_x-\frac{3}{2}\Omega\Delta B_x\ ,
\end{equation}
where $\eta_{Hz}\propto B_z$ is the Hall diffusivity based on vertical magnetic field,
assumed to be constant to facilitate the analysis, and $\Delta B_x$ represents changes
in $B_x$ to account for additional shear conversion of $B_x$ to $-B_y$.

The first equation (\ref{eq:hallshr1}) describes the generation of radial field due to the
Hall effect. The underlying physics is best understood from the grain-free expression of
$\eta_H$ in Equation (\ref{eq:diff0}): Vertical gradient of toroidal field provides radial
current $\pa B_y/\pa z=-J_x\propto v_{e,x}-v_{i,x}$, corresponding to radial drift of
electrons relative to the ions. In the Hall dominated regime, the ions are coupled to the
neutrals $v_{i,x}\sim v_x$, while magnetic field is frozen to the electrons. The second
$z$-derivative then descries conversion of vertical field into radial fields due to
vertical shear of electron motion. Depending on the sign of $B_z$ (hence
$\eta_{Hz}$), the result is that the original radial field can be amplified ($B_z>0$) or
reduced ($B_z<0$) once the Hall term is turned on, as we see in Figure \ref{fig:fiducial}.

The second equation (\ref{eq:hallshr2}) provides positive feedback to field evolution due
to shear. For $B_z>0$, shear conversion from $\Delta B_x$ amplifies the toroidal field
(and in general its second derivative), which promotes additional amplification of the radial
field via (\ref{eq:hallshr1}), leading to runaway. This is closely related to the Hall-shear
instability of \citet{Kunz08}, as also pointed out in \citet{Lesur_etal14}. We see from
Figure \ref{fig:fiducial} that the radial field $B_x$ is substantially stronger than the Hall-free
case throughout the disk interior. Similarly, the toroidal field also becomes much stronger
than the Hall-free case. The field amplification process is eventually saturated due to
damping by Ohmic resistivity and AD, as well as advection of magnetic field by disk outflow.

The opposite applies when $B_z<0$. The Hall effect and shear act destructively to the
Hall-free field configuration and both radial and toroidal magnetic fields are reduced. In
particular, we see from the left panels of Figure \ref{fig:fiducial} that the radial field
even changes sign around $z=3H$ due to the Hall effect, hence $B_x$ and $B_y$ have
the same sign (but small amplitude) around this region, giving a negative Maxwell stress.
While shear conversion tends to reverse the sign of $B_y$ as well, this does not occur
due to AD.

\subsection[]{Issues with Symmetry}\label{ssec:issue}

While we have enforced reflection symmetry across the disk midplane to guarantee
that the wind solutions have physical geometry, it remains to clarify to what extent
this assumption can be justified. In particular, in the case of $B_z>0$, this treatment
forces $B_x$ and $B_y$ to zero at the midplane, which works against magnetic
field amplification. To this end, we perform a set of additional simulations containing
the full disk with fiducial parameters (1AU, $\beta_0=10^5$ with two polarities),
named R1b5H$*$Full.

In these simulations, we follow a similar procedure by starting with a Hall-free
run to $t=480\Omega^{-1}$, and then turn on the Hall effect for the two polarities.
The Hall-free run saturates into the odd-$z$ symmetry solution (unphysical for
the wind), where the horizontal magnetic field maximizes at the midplane. After
turning on the Hall effect, the same symmetry remains, which we run to time
$t=600\Omega^{-1}$ for full relaxation. At this time, we {\it manually} flip the
horizontal magnetic field and velocity field at all cells with $z<0$ to achieve the
physical even-$z$ symmetry. We then continue to run the simulations to
$t=1080\Omega^{-1}$ and focus on how the system relaxes. We call these two
continued runs R1b5H$\pm$Full. For comparison, we also apply the flip to the
Hall-free simulation, which is named as R1b5H0Full.

\begin{figure*}
    \centering
    \includegraphics[width=160mm]{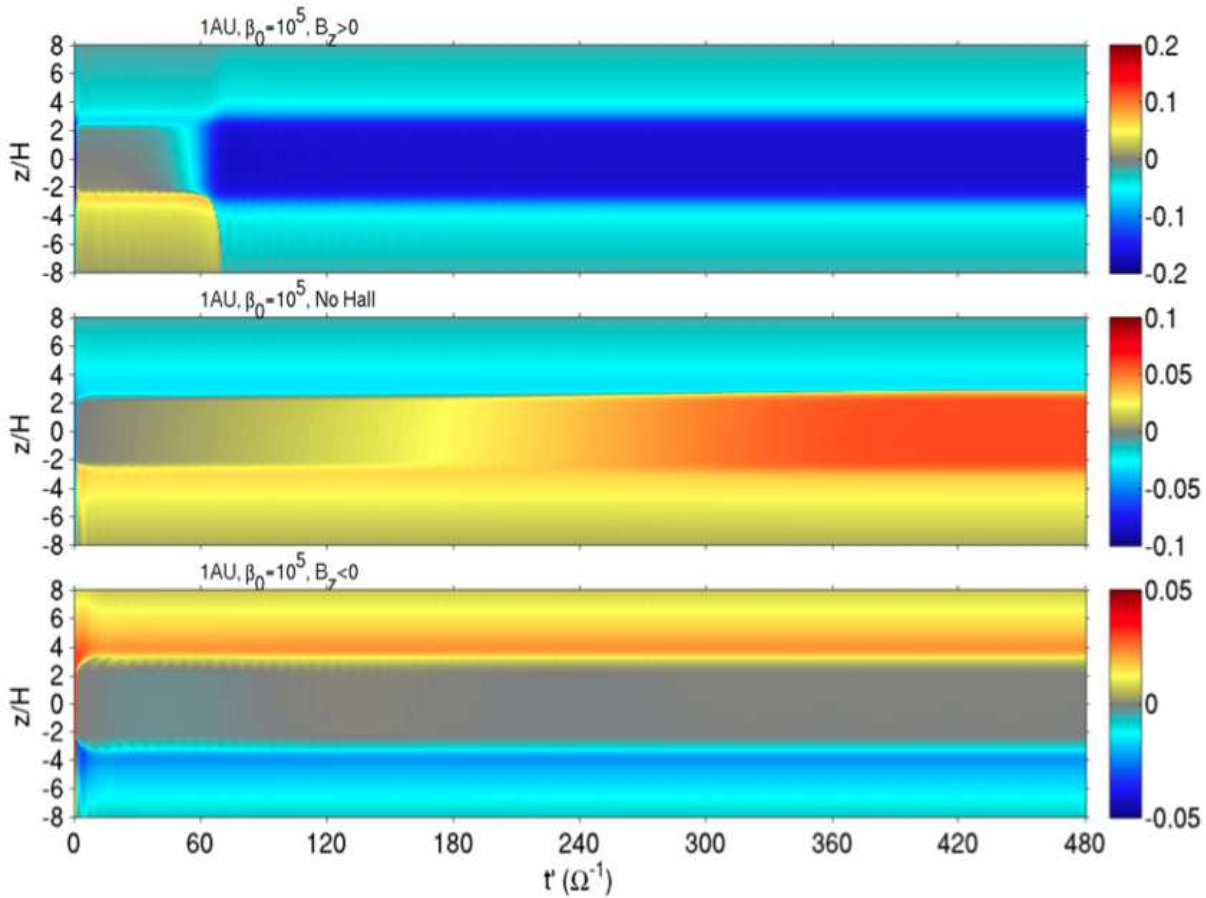}
  \caption{Time evolution of the magnetic field profile $B_y$ in our full-disk simulations
  R1b5H$*$Full after a manual flip is performed to achieve even-$z$ symmetry (see
  Section \ref{ssec:issue}). The top, middle bottom panels correspond to $B_z>0$,
  Hall-free and $B_z<0$ cases.}\label{fig:relax-full}
\end{figure*}

In Figure \ref{fig:relax-full}, we show the time evolution of the toroidal field
profiles from these the runs after the flip (we set $t'=0$ at the time of flip).
In the Hall-free case, we see that the system maintains the even-$z$ symmetry
state for a few orbits, while asymmetry slowly develops and the midplane toroidal
field is gradually amplified with a single sign. Nevertheless, the magnetic field
configuration in the wind zone (surface layer) remains unchanged, and the system
eventually relaxes to the physical wind solution with a strong current layer offset
from disk midplane at $z\sim3H$, as highlighted in \citet{BaiStone13b}.

\subsubsection[]{The $B_z>0$ Case}

With the Hall effect, and when $B_z>0$, we see that the initial evolution of the
system is similar to the Hall-free case, but in 10 orbits, the midplane field gets rapidly
amplified (as a result of the Hall-shear instability), and the horizontal field flips back
to arrive at the odd-$z$ symmetry solution. There is a transient phase (around
$t'\sim60\Omega^{-1}$) where the field configuration remains physical for a disk
wind and contains a strong current layer at $z\sim-2H$, but the field amplification is so
rapid that a unidirectional toroidal field quickly overwhelms and spreads into the entire
disk. Overall, it appears that with strong field amplification due to the Hall-shear instability,
the even-$z$ symmetry solution is difficult to be maintained in shearing-box without
manually enforcing the symmetry at the midplane\footnote{However, this can be achieved
at outer disk radii, as we will demonstrate in the forthcoming paper with an example at 5 AU.}.
In addition, achieving a physical wind solution with a strong current layer offset from the
midplane appears difficult as well. This point was also raised in \citet{Lesur_etal14} based
on their simulations.

The above fact poses serious concerns about the physical reality of the system at $\sim1$
AU: one either achieves the odd-$z$ symmetry solution with unphysical wind geometry, or
achieves the more physical even-$z$ symmetry solution by unrealistically restricting the field
geometry. This apparent dilemma may reflect the limitations of the shearing-box framework,
and global simulations will be the key to resolving these issues. While we choose to restrict
the symmetry in our simulations for most of this work, readers should bare in mind the
potential caveats.

\subsubsection[]{The $B_z<0$ Case}

\begin{figure}
    \centering
    \includegraphics[width=90mm]{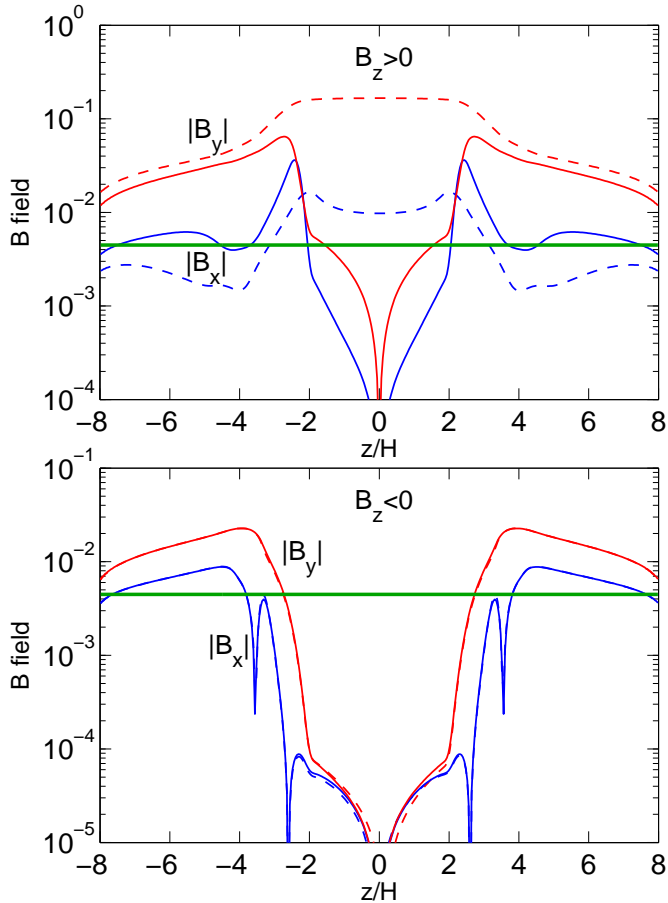}
  \caption{Time evolution of the magnetic field profile $B_y$ in our full-disk simulations
  R1b5H$\pm$Full after even-$z$ symmetry is enforced (see Section \ref{ssec:issue}).
  The top, middle and bottom panels correspond to $B_z>0$, Hall-free and $B_z<0$ runs.
  }\label{fig:symmetry}
\end{figure}

When $B_z<0$, on the other hand, we see from the bottom panel of Figure
\ref{fig:relax-full} that the physical even-$z$ symmetry solution we obtained earlier easily
survives in the full-disk simulation. The bottom panel of Figure \ref{fig:symmetry} further
compares the full-disk solution with our previous solution with enforced reflection
symmetry. The agreement is almost exact. Therefore, we conclude that for the $B_z<0$
case, a wind solution with physical geometry exists naturally, where the horizontal field
diminishes around the midplane region due to the Hall effect and transitions through
zero.

\subsection[]{General Properties of the New Wind Solutions}

As introduced in Section \ref{ssec:symmetry}, we separate the wind solution into
a disk zone and a wind zone. They are divided at $z=z_b$, the base of the wind.
Conventionally, $z_b$ is defined as the point where the azimuthal velocity
transitions from sub-Keplerian to super-Keplerian \citep{WardleKoenigl93}, which
is adopted in our previous studies (i.e., $v_y=0$ in shearing-box simulations). For
our new wind solutions, we find that this location is well defined when $B_z<0$,
as can be seen from the second row of Figure \ref{fig:fiducial}, where $|v_y|$ shows
a clear kink in the logarithmic plot at about $z=4.2H$. This is only slightly smaller
than $z_b\approx4.3H$ in the Hall-free run R1b5H0. When $B_z>0$, we see that
at about the same location, $v_y$ undergoes a minimum but does not
reverse sign. Since other aspects of this solution does not change significantly,
we modify the definition of $z_b$ as follows: moving from disk surface downward,
$z_b$ is located at where  $|v_y|$ experiences a minimum for the first time. This
definition maintains consistency with our Hall-free run R1b5H0. Also, for run R1b5H+,
the Reynolds stress $\rho v_zv_y$ is minimized at $z_b$ and is negligible compared
with the Maxwell stress $(-B_yB_z)$.

Below we discuss the general properties of the new wind solutions, focusing on
the fiducial (half-disk) runs R1b5H$\pm$. For the $B_z>0$ case, we further
compare the wind properties between half and full-disk simulations at the end
of this subsection.

\subsubsection[]{Angular Momentum Transport by Disk Wind}

Angular momentum transport by disk wind has been discussed in Section
\ref{ssec:symmetry}. With the physical wind symmetry as enforced in our simulations,
the wind-driven accretion rate $\dot{M}_{\rm V}$ is proportional to the wind stress
$T_{z\phi}^{z_b}$, and can be estimated by Equation (\ref{eq:dotMV}). Their values
are provided in Table \ref{tab:runs1}. For our fiducial runs R1b5H$\pm$,
the wind-driven accretion rates in both magnetic polarities are well above the
desired value of $10^{-8}M_{\bigodot}$ yr$^{-1}$.

We see that including the Hall term, the wind-driven accretion rates are modestly
increased (reduced) in the case of $B_z>0$ ($B_z<0$) compared with the
Hall-free run R1b5H0. Since $T_{z\phi}=-B_yB_z$ with $B_z$ being
constant in the disk, the modest increase/reduction is directly related to the
amplification/reduction of $B_y$ discussed in Section \ref{ssec:relax}.

\subsubsection[]{Radial Transport of Angular Momentum}

\begin{figure}
    \centering
    \includegraphics[width=90mm]{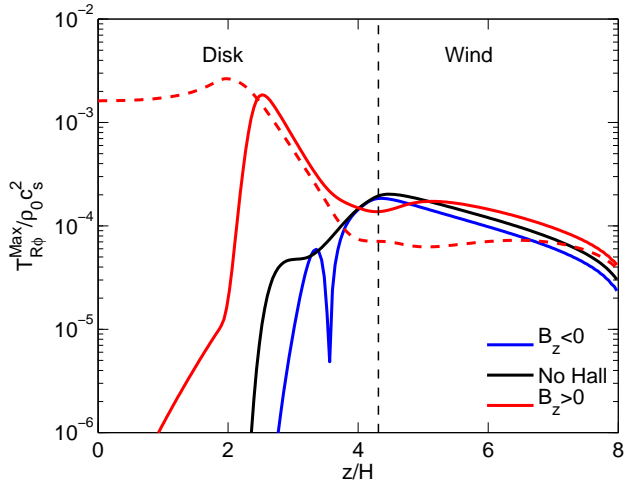}
  \caption{The vertical profile of the Maxwell stress $T_{R\phi}^{\rm Max}$ for
  our three fiducial simulations R1b5H$-$ (blue), R1b5H0 (black) and R1b5H+
  (red). The red dashed line corresponds to run R1b5H+Full.
  Note that in all cases the sign of $T_{R\phi}^{\rm Max}$ is positive except
  the left part of the blue curve. The vertical dashed line marks the location of
  $z_b$, the base of the wind.}\label{fig:maxwell}
\end{figure}

Radial transport of angular momentum via magnetic braking has been discussed
in Section \ref{ssec:dLinR}. In Figure \ref{fig:maxwell} we show the vertical profiles
of $T_{R\phi}^{\rm Max}$ from our fiducial runs. In Table \ref{tab:runs1} we further
list the value of $\alpha^{\rm Max}$, and the corresponding $\dot{M}_{\rm R}$
assuming an MMSN disk. We see that including the Hall term, $\alpha^{\rm Max}$ is
substantially enhanced (reduced) in the case of $B_z>0$ ($B_z<0$). For
$B_z>0$, the enhancement of $T_{R\phi}^{\rm max}$ is greatest in the
Hall-dominated region ($z\lesssim3H$). This mainly results from magnetic field
amplification as discussed in Section \ref{ssec:relax}. We see from Figure
\ref{fig:fiducial} that $B_x$ in this region is amplified by more than an order of
magnitude compared with the Hall-free case. Together with modestly amplified
$B_y$, they both contribute to enhance the total $\alpha^{\rm Max}$ by a factor
of $\sim8$ compared with the Hall-free case. For $B_z<0$, the reduction and
reversal of $B_x$ together with reduced $B_y$ naturally leads to much smaller
$\alpha^{\rm Max}$.

The radial transport of angular momentum via magnetic braking was mentioned but
not emphasized in our earlier work of \citet{BaiStone13b}, since its contribution is much
smaller than that from wind-driven accretion. With the Hall effect and aligned magnetic
field, however, magnetic braking already contributes a non-negligible
fraction of the angular momentum transport, as read from Table \ref{tab:runs1}, and
this term alone is sufficient to account for the typically observed accretion rates in PPDs.
On the other hand, for anti-aligned magnetic field geometry, contribution from magnetic
braking is completely negligible.

\begin{table*}
\caption{Fiducial Runs and Variations.}\label{tab:runs1}
\begin{center}
\begin{tabular}{c|cc|cc|c|cc|c|cc}\hline\hline
 Run  & $\alpha^{\rm Max}$ & $\dot{M}_{{\rm R},-8}$ & $T_{z\phi}^{z_b}$ &
 $\dot{M}_{{\rm V},-8}$ & $\dot{M}_w$ 
 & $v_{\rm in,max}$ & $z_{\rm in, max}$ &$v_{Bx}$ & $z_b$ & $z_A$ \\\hline
R1b5H+ & $1.08\times10^{-3}$ & 0.89 & $1.53\times10^{-4}$ & $6.27$ & $2.81\times10^{-5}$
& $-0.023$ & $2.48$ & $-0.24$ & $4.31$ & $5.06$\\
R1b5H0 & $1.23\times10^{-4}$ & 0.10 & $1.19\times10^{-4}$ & $4.87$ & $1.65\times10^{-5}$
& $-0.012$ & $2.56$ & $-1.0\times10^{-3}$ & $4.31$ & $5.98$\\
R1b5H-- & $5.25\times10^{-5}$ & $0.043$ & $9.75\times10^{-5}$ & $4.00$ & $1.19\times10^{-5}$
& $-0.072$ & $3.48$ & $0.074$ & $4.27$ & $6.40$\\\hline

R1b5H+Full & $4.48\times10^{-3}$ & 3.67 & $1.80\times10^{-4}$ & $7.41$ & $4.90\times10^{-5}$
& -- & -- & $\sim0$ & $4.69$ & $4.79$\\
R1b5H0Full & $2.20\times10^{-4}$ & 0.18 & $1.19\times10^{-4}$ & $4.87$ & $1.69\times10^{-5}$
& $-0.29$ & $2.90$ & $\sim0$ & $4.39$ & $5.92$\\
R1b5H--Full & $5.92\times10^{-5}$ & $0.043$ & $9.66\times10^{-5}$ & $3.96$ & $1.21\times10^{-5}$
& $-0.099$ & $3.44$ & $0.073$ & $4.27$ & $6.40$\\\hline

M03-R1b5H+ & $4.36\times10^{-3}$ & $1.07$ & $5.12\times10^{-4}$ & $6.30$ & $9.30\times10^{-5}$
& $-0.024$ & $1.94$ & $-0.22$ & $4.06$ & $4.81$\\
M03-R1b5H0 & $4.26\times10^{-4}$ & $0.10$ & $3.93\times10^{-4}$ & $4.84$ & $5.34\times10^{-5}$
& $-0.011$ & $2.06$ & $-7.5\times10^{-4}$ & $4.06$ & $5.81$\\
M03-R1b5H-- & $1.90\times10^{-4}$ & $0.047$ & $3.23\times10^{-4}$ & $3.97$ &$3.86\times10^{-5}$
& $-0.073$ & $3.15$ & $0.067$ & $4.02$ & $6.27$\\\hline

M3-R1b5H+ & $3.11\times10^{-4}$ & $0.77$ & $5.10\times10^{-5}$ & $6.27$ & $9.48\times10^{-6}$
& $-0.022$ & $2.90$ & $-0.25$ & $4.52$ & $5.23$\\
M3-R1b5H0 & $4.12\times10^{-5}$ & $0.10$ & $3.95\times10^{-5}$ & $4.86$ &$5.65\times10^{-6}$
& $-0.012$ & $2.98$ & $-1.1\times10^{-3}$ & $4.56$ & $6.10$\\
M3-R1b5H-- & $1.80\times10^{-5}$ & $0.044$ & $3.27\times10^{-5}$ & $4.02$ &$4.04\times10^{-6}$
& $-0.071$ & $3.73$ & $0.080$ & $4.52$ & $6.56$\\\hline

nogr-R1b5H+ & $1.05\times10^{-2}$ & $8.62$ & $2.42\times10^{-4}$ & $9.94$ & $4.63\times10^{-5}$
& $-2.1\times10^{-3}$ & $1.19$ & $-0.28$ & $4.06$ & $4.98$\\
nogr-R1b5H+Full & $1.40\times10^{-2}$ & $8.62$ & $2.42\times10^{-4}$ & $9.94$ & $4.63\times10^{-5}$
& $-2.1\times10^{-3}$ & $1.19$ & $-0.28$ & $4.06$ & $4.98$\\
nogr-R1b5H0  & $2.23\times10^{-4}$ & $0.18$ & $1.29\times10^{-4}$ & $5.29$ & $1.91\times10^{-5}$
& $-7.9\times10^{-3}$ & $2.31$ & $6.8\times10^{-4}$ & $4.40$ & $5.81$\\
nogr-R1b5H-- & $7.67\times10^{-5}$ & $0.063$ & $1.11\times10^{-4}$ & $4.57$ & $1.47\times10^{-5}$
& $-0.048$ & $3.27$ & $0.022$ & $4.27$ & $6.10$\\\hline

X29-R1b5H+ & $4.90\times10^{-4}$ & 0.40 & $1.46\times10^{-4}$ & $6.02$ & $2.58\times10^{-5}$
& $-0.022$ & $2.81$ & $-0.44$ & $4.10$ & $5.02$\\
X29-R1b5H0 & $1.28\times10^{-4}$ & 0.11 & $1.15\times10^{-4}$ & $4.72$ & $1.57\times10^{-5}$
& $-0.012$ & $2.69$ & $0.019$ & $4.31$ & $6.02$\\\hline

X31-R1b5H+ & $1.84\times10^{-3}$ & 1.51 & $2.43\times10^{-4}$ & $9.97$ & $3.19\times10^{-5}$
& $-0.026$ & $2.27$ & $-0.075$ & $3.73$ & $5.31$\\
X31-R1b5H0 & $1.73\times10^{-4}$ & 0.15 & $1.55\times10^{-4}$ & $6.35$ & $2.01\times10^{-5}$
& $-0.014$ & $2.40$ & $-4.6\times10^{-4}$ & $3.94$ & $5.77$\\
X31-R1b5H--$^*$ & $7.16\times10^{-5}$ & 0.059 & $1.18\times10^{-4}$ & $4.85$ & $1.52\times10^{-5}$
& $-0.14$ & $3.31$ & $-1.3\times10^{-3}$ & $4.10$ & $6.02$\\\hline

\hline\hline
\end{tabular}
\end{center}
See Section 3.2 for description of simulation runs and naming conventions. The last run (with $^*$) is
eventually unstable, where values are taken before the instability takes over. The results are mainly
discussed in Section 5.1. 

List of physical quantities in the Table are, $\alpha^{\rm Max}$:
Shakura-Sunyaev $\alpha$ due to Maxwell stress (magnetic braking);
$\dot{M}_{{\rm R},-8}$: accretion rate due to radial transport of angular momentum
($10^{-8}M_{\bigodot}$ yr$^{-1}$); $T_{z\phi}^{z_b}$: the wind stress (natural unit);
$\dot{M}_{{\rm V},-8}$: wind-driven accretion rate ($10^{-8}M_{\bigodot}$ yr$^{-1}$);
$\dot{M}_w$: single-sided mass outflow rate ($\rho_0c_s$),
$v_{\rm in,max}$: maximum inflow velocity ($c_s$);
$z_{\rm in, max}$: location at the maximum inflow velocity ($H$),
$v_{Bx}$: radial drift velocity of vertical magnetic flux;
$z_b$: location of the base of the wind;
$z_A$:  location of the Alfv\'en point.
\end{table*}

\begin{table*}
\caption{List of simulations for extended parameter study.}\label{tab:runs2}
\begin{center}
\begin{tabular}{c|cc|cc|c|cc|c|cc}\hline\hline
 Run  & $\alpha^{\rm Max}$ & $\dot{M}_{{\rm R},-8}$ & $T_{z\phi}^{z_b}$ &
 $\dot{M}_{{\rm V},-8}$ & $\dot{M}_w$ 
 & $v_{\rm in,max}$ & $z_{\rm in, max}$ &$v_{Bx}$ & $z_b$ & $z_A$ \\\hline

R03b5H+ & $2.71\times10^{-4}$ & $0.41$ & $1.00\times10^{-4}$ & $10.1$ & $9.89\times10^{-6}$
& $-0.072$ & $3.27$ & $-0.36$ & $4.06$ & $6.19$\\
R03b5H0 & $6.86\times10^{-5}$ & $0.10$ & $7.20\times10^{-5}$ & $7.28$ & $6.90\times10^{-6}$
& $-4.9\times10^{-2}$ & $3.19$ & $-1.0\times10^{-3}$ & $4.23$ & $7.06$\\
R03b5H- & $3.26\times10^{-5}$ & $0.049$ & $5.52\times10^{-5}$ & $5.58$ & $5.49\times10^{-6}$
& $-0.12$ & $3.69$ & $0.17$ & $4.44$ & $7.48$\\\hline

R03b6H+ & $1.29\times10^{-4}$ & $0.19$ & $1.94\times10^{-5}$ & $1.96$ & $4.15\times10^{-6}$
& $-0.015$ & $3.15$ & $-0.19$ & $4.69$ & $5.23$\\
R03b6H0 & $1.39\times10^{-5}$ & $0.021$ & $1.48\times10^{-5}$ & $1.49$ & $2.33\times10^{-6}$
& $-1.3\times10^{-2}$ & $3.35$ & $-1.7\times10^{-3}$ & $4.69$ & $5.98$\\\hline

R1b4H+ & $1.81\times10^{-3}$ & $1.49$ & $7.37\times10^{-4}$ & $30.2$ & $6.11\times10^{-5}$
& $-0.096$ & $2.69$ & $-0.40$ & $3.60$ & $6.60$\\
R1b4H0 & $6.13\times10^{-4}$ & $0.50$ & $5.51\times10^{-4}$ & $22.6$ & $4.64\times10^{-5}$
& $-0.047$ & $2.44$ & $-3.6\times10^{-3}$ & $3.81$ & $7.40$\\
R1b4H-- & $3.03\times10^{-4}$ & $0.25$ & $4.39\times10^{-4}$ & $18.0$ & $3.90\times10^{-5}$
& $-0.13$ & $3.02$ & $0.20$ & $3.98$ & $7.77$\\\hline
 
R1b5H+ & $1.08\times10^{-3}$ & 0.89 & $1.53\times10^{-4}$ & $6.27$ & $2.81\times10^{-5}$
& $-0.023$ & $2.48$ & $-0.24$ & $4.31$ & $5.06$\\
R1b5H0 & $1.23\times10^{-4}$ & 0.10 & $1.19\times10^{-4}$ & $4.87$ & $1.65\times10^{-5}$
& $-0.012$ & $2.56$ & $-1.0\times10^{-3}$ & $4.31$ & $5.98$\\
R1b5H-- & $5.25\times10^{-5}$ & $0.043$ & $9.75\times10^{-5}$ & $4.00$ & $1.19\times10^{-5}$
& $-0.072$ & $3.48$ & $0.074$ & $4.27$ & $6.40$\\\hline

R1b6H+ & $3.32\times10^{-4}$ & $0.27$ & $3.97\times10^{-5}$ & $1.63$ & $1.06\times10^{-5}$
& $-4.4\times10^{-3}$ & $2.40$ & $-0.12$ & $4.19$ & $4.64$\\
R1b6H0 & $1.77\times10^{-5}$ & $0.015$ & $2.57\times10^{-5}$ & $1.05$ & $4.81\times10^{-6}$
& $-3.1\times10^{-3}$ & $2.73$ & $-7.4\times10^{-4}$ & $4.40$ & $5.44$\\\hline

R3b4H+ & $1.29\times10^{-2}$ & $6.11$ & $1.15\times10^{-3}$ & $20.7$ & $1.77\times10^{-4}$
& $-1.6\times10^{-2}$ & $1.52$ & $-0.27$ & $3.98$ & $4.94$\\
R3b4H0 & $1.07\times10^{-3}$ & $0.51$ & $9.00\times10^{-4}$ & $16.2$ & $1.07\times10^{-4}$
& $-8.7\times10^{-3}$ & $1.73$ & $-2.1\times10^{-4}$ & $3.94$ & $6.15$\\
R3b4H-- & $5.73\times10^{-4}$ & $0.27$ & $7.98\times10^{-4}$ & $14.3$ & $8.68\times10^{-5}$
& $-0.052$ & $2.64$ & $0.11$ & $3.90$ & $6.52$\\\hline

R3b5H+ & $4.79\times10^{-3}$ & $2.27$ & $2.89\times10^{-4}$ & $5.21$ & $6.79\times10^{-5}$
& $-1.8\times10^{-3}$ & $1.40$ & $-0.093$ & $3.85$ & $4.39$\\
R3b5H0 & $1.66\times10^{-4}$ & $0.078$ & $2.04\times10^{-4}$ & $3.67$ & $3.41\times10^{-5}$
& $-2.5\times10^{-3}$ & $1.85$ & $-1.1\times10^{-4}$ & $4.02$ & $5.27$\\\hline

R3b6H+ & $1.23\times10^{-3}$ & $0.58$ & $7.76\times10^{-5}$ & $1.40$ & $2.39\times10^{-5}$
& $-4.5\times10^{-4}$ & $1.44$ & $-0.058$ & $3.85$ & $4.19$\\\hline

R5b4H+ & $2.06\times10^{-2}$ & $7.54$ & $1.52\times10^{-3}$ & $18.6$ & $2.68\times10^{-4}$
& $-0.017$ & $0.23$ & $-0.099$ & $3.81$ & $4.56$\\
R5b4H0 & $1.28\times10^{-3}$ & $0.47$ & $1.17\times10^{-3}$ & $14.4$ & $1.55\times10^{-4}$
& $-4.0\times10^{-3}$ & $0.0$ & $-1.1\times10^{-4}$ & $3.81$ & $5.64$\\
R5b4H-- & $6.89\times10^{-4}$ & $0.25$ & $1.02\times10^{-3}$ & $12.5$ & $1.22\times10^{-4}$
& $-0.029$ & $2.40$ & $0.073$ & $3.77$ & $6.02$\\\hline

R5b5H+ & $6.29\times10^{-3}$ & $2.31$ & $3.95\times10^{-4}$ & $4.84$ & $1.02\times10^{-4}$
& $-5.0\times10^{-3}$ & $0.10$ & $-0.046$ & $3.69$ & $4.14$\\
R5b5H0 & $1.86\times10^{-4}$ & $0.068$ & $2.54\times10^{-4}$ & $3.11$ & $4.50\times10^{-5}$
& $-6.9\times10^{-4}$ & $1.52$ & $-7.2\times10^{-5}$ & $3.90$ & $5.06$\\\hline

R5b6H+ & $2.13\times10^{-3}$ & $0.78$ & $9.00\times10^{-5}$ & $1.10$ & $3.86\times10^{-5}$
& $-1.3\times10^{-3}$ & $0.0$ & $-0.043$ & $4.64$ & $3.94$\\\hline

R8b4H+ & $1.23\times10^{-2}$ & $3.57$ & $1.81\times10^{-3}$ & $15.6$ & $3.12\times10^{-4}$
& $-0.028$ & $0.10$ & $-6.5\times10^{-3}$ & $3.73$ & $4.73$ \\
R8b4H0 & $1.84\times10^{-3}$ & $0.53$ & $1.52\times10^{-3}$ & $13.1$ & $2.20\times10^{-4}$
& $-1.1\times10^{-2}$ & $0.0$ & $0$ & $3.65$ & $5.27$\\\hline

R8b5H+ & $3.65\times10^{-3}$ & $1.06$ & $4.92\times10^{-4}$ & $4.24$ & $1.13\times10^{-4}$
& $-7.6\times10^{-3}$ & $0.06$ & $-3.8\times10^{-3}$ & $3.56$ & $4.27$\\\hline

R8b6H+ & $1.16\times10^{-3}$ & $0.34$ & $1.37\times10^{-4}$ & $1.18$ & $3.74\times10^{-5}$
& $-1.9\times10^{-3}$ & $0.0$ & $-3.7\times10^{-3}$ & $3.52$ & $4.10$\\\hline

R15b4H+ & $6.72\times10^{-3}$ & $1.42$ & $2.30\times10^{-3}$ & $12.4$ & $4.01\times10^{-4}$
& $-0.027$ & $0.06$ & $-1.6\times10^{-3}$ & $3.52$ & $4.73$ \\

\hline\hline
\end{tabular}
\end{center}
Same as Table \ref{tab:runs1}, and see Section 3.2 for description of simulation runs and naming
conventions. Results are mainly discussed in Section 5.3-5.4.
\end{table*}

\subsubsection[]{Wind-driven Accretion Flow}

In our simulations, the wind stress $T_{z\phi}$ directly leads to an inward
accretion mass flux, while there is no mass flux associated with radial angular
momentum transport due to the shearing-sheet formulation where radial gradients
are ignored. Here we focus on the wind-driven accretion mass flux.

We see from the second row of Figure \ref{fig:fiducial} that in all three runs
R1b5H$(\pm,0)$, the radial velocity transitions from being positive in the wind
zone, to negative somewhere below the base of the wind, which corresponds to
the accretion flow. Interestingly, the vertical distribution of the accretion flow is
different in the three runs: for $B_z<0$, the inflow region is located at larger
vertical hight compared with the Hall-free case, while for $B_z>0$, the inflow is
located further toward disk interior. To characterize the basic properties of such
inward mass flux, we identify the maximum inflow velocity $v_{\rm in, max}$ and
the location where it is achieved $z_{\rm in, max}$, and list their values in Table
\ref{tab:runs1}.

We note that the location of the inflow corresponds to where the wind stress
$T_{z\phi}$ is exerted to the disk. The vertical gradient of $T_{z\phi}$ (or
effectively $B_y$) is directly related to the torque per unit length received by
the gas hence the rate of the gas inflow. When $B_z>0$, the horizontal magnetic
field is amplified toward disk interior, therefore, the inflow region is located closer
to the disk midplane. This can be effectively interpreted as that $B_z>0$ allows
the magnetic field to be coupled with the gas deeper toward the midplane.
The opposite applies for the $B_z<0$ case.

\subsubsection[]{Disk Outflow}

The outflow mass loss rate in the disk wind is not well characterized in shearing-box
simulations. It decreases with increasing the vertical box size, as studied and
discussed extensively in \citet{Fromang_etal13} for the MRI turbulence case and
\citet{BaiStone13b} for the laminar wind case. Here, we are not concerned with the
absolute mass loss rate, but focus on the relative dependence of the mass outflow
rate on physical parameters such as magnetic polarity (this subsection), $\beta_0$,
and disk radius (Section \ref{sec:param}), where shearing-box may provide more
reliable results.

The measured mass loss rates $\rho v_z$ from our simulations are listed in Table
\ref{tab:runs1}.\footnote{Note that the values reported correspond to single-sided mass
loss rate, while the Tables in \citet{BaiStone13b} and \citet{Bai13} quote the mass loss
rates from both sides of the disk.} We see that when $B_z>0$ ($B_z<0$), the wind
mass loss rate is higher (lower) compared with the Hall-free case, consistent with the
wind being stronger (weaker) discussed earlier. The increase (reduction) in the outflow
mass flux is mainly due to the higher (lower) gas density at $z=z_b$, as a result of
stronger (weaker) magnetic pressure support. The change in the mass outflow rate is
accompanied by the change in the location of the Alfv\'en point, $z_A$. It is defined as
the location where vertical velocity equals to the vertical Alfv\'en velocity
$v_z(z_A)^2=B_z^2/4\pi\rho(z_A)$. As discussed in \citet{BaiStone13b}, larger outflow
rate makes the Alfv\'enic point lower, and vice versa (see their Section 4.5).

\subsubsection[]{Magnetic Flux Transport}

Poloidal magnetic flux can drift radially in the disk at velocity $v_{Bx}$ in the
presence of toroidal electric field $E_y$
\begin{equation}
v_{Bx}=-\frac{E_y}{B_z}\ ,
\end{equation}
where ${\mb E}$ is given in Equation (\ref{eq:emf}). The steady state condition
further requires $E_y$ to be constant with height so that magnetic flux drifts
uniformly across the disk, giving a single value of $v_{Bx}$. Positive or negative
$v_{Bx}$ would lead to expulsion or accumulation of magnetic flux.

In Table \ref{tab:runs1}, we show the value of $v_{Bx}$ measured from our
simulations. We see that without the Hall term, the value of $v_{Bx}$ is very close
to zero in run R1b5H0, as found earlier in \citet{BaiStone13b}. Including the Hall
term, $v_{Bx}$ deviates substantially from $0$, and is negative (positive) when
$B_z>0$ ($B_z<0$). This means that poloidal magnetic flux is transported
inward (outward) at large velocities ($5\%-15\%c_s$), much faster than the velocity
of the accretion flow.

In reality, the value of $v_{Bx}$ should be determined by global conditions and can
not be controlled in our local simulations. Therefore, we may expect that the realistic
value of $v_{Bx}$ should be much closer to zero than what we obtain here. The
general properties of the wind solution have been found to depend weakly on the
exact value of $v_{Bx}$ \citep{WardleKoenigl93}. Moreover, the measured values of
$v_{Bx}$ are still much less than the sound speed, hence we do not expect the
properties of the supersonic wind to be strongly affected. Overall, the values of
$v_{Bx}$ listed in Table \ref{tab:runs1} should mainly be taken for reference but not
to be taken seriously for studying magnetic flux transport.

\subsubsection[]{Comparison with Full-disk Simulations}

Finally, we compare our fiducial half-disk simulations R1b5H$*$ with enforced
reflection symmetry with full-disk simulations R1b5H$*$Full. As discussed in Section
\ref{ssec:issue}, for $B_z<0$, the full-disk simulation yields almost exactly
the same wind solution as half-disk simulations. Table \ref{tab:runs1} further
confirms that major diagnostics between runs R1b5H$-$ and R1b5H$-$Full are
almost identical. In the Hall-free case, the wind diagnostics between runs
R1b5H0 and R1b5H0Full are also very close, with the full-disk run producing
higher $\alpha^{\rm Max}$, consistent with the results in \citet{BaiStone13b}.

Below we focus on the comparison for the $B_z>0$ case. The top panel of Figure
\ref{fig:symmetry} compares the magnetic field profiles between the odd-$z$ and
even-$z$ symmetry solutions. With a full disk, we see that the horizontal magnetic
fields $B_x$ and $B_y$ get amplified and maintain its strength across the midplane,
instead of being forced to damp to zero within $z=\pm2H$ in the half-disk run.
As a result, large $T_{R\phi}^{\rm Max}$ extends to the midplane, leading to
stronger magnetic braking. Reading from Table \ref{tab:runs1}, we see that
$\alpha^{\rm Max}$ in run R1b5H+Full is about 4 times higher than in run R1b5H+.

We also notice that at the disk surface, the odd-$z$ and even-$z$ symmetry
solutions do not overlap. This is different from the Hall-free case, where they match
each other at the disk surface \citep{BaiStone13b}. The reason is that odd-$z$
symmetry solution (in the full-disk run) requires $E_y=0$ by construction, thus
$v_{Bx}=0$; while we have seen the even-$z$ symmetry solutions from the half-disk
runs have non-zero $v_{Bx}$. Therefore, the odd-$z$ symmetry solutions we have
obtained are not even-$z$ symmetry solutions with flipped horizontal field. Major
wind diagnostics listed in Table \ref{tab:runs1} show that both $T_{z\phi}^{z_b}$ and
$\dot{M}_w$ are larger in full-disk simulations by about $20\%$ and $75\%$
respectively.

\section[]{Simulation Results: Parameter Study of the Wind Solutions}\label{sec:param}

In this section, we consider much wider range of parameters and discuss how they affect the
properties of the new disk wind solutions. We first consider variations to our fiducial solution
at 1AU with $\beta_0=10^5$ in Section \ref{ssec:vari}, with the list of runs and results shown in
Table \ref{tab:runs1}. We then vary the radial location and vertical field strength and discuss
the results in Sections \ref{ssec:stable} to \ref{ssec:scale}, with the list of runs provided in Table
\ref{tab:runs2}. Note that when varying $R_{\rm AU}$, the ionization profile changes which
changes the {\it absolute} strength of all non-ideal MHD terms simultaneously, meanwhile,
changes in gas density alters the {\it relative} importance among the three non-ideal MHD effects,
as can be inferred from Equation (\ref{eq:diff0}). Correspondingly, Ohmic resistivity becomes
progressively less important toward large radii, where AD becomes progressively more
prevailing.

\subsection[]{Variations to Fiducial Runs}\label{ssec:vari}

\subsubsection[]{Disk Surface Density}\label{sssec:sigma}

We first vary the disk surface density by a factor of $3$ and $0.3$ (labeled by ``M3" and ``M03").
Accordingly, we have also varied the strength of the net vertical magnetic field so that the
physical value of the field strength remains the same (hence $\beta_0=3\times10^5$ and
$3\times10^4$ respectively). Looking from Table \ref{tab:runs1} we see that the
wind-driven accretion rate $\dot{M}_V$ almost remain identical as in the fiducial run under
these variations, for both $B_z>0$ and $B_z<0$. Similarly, the measured wind mass loss rate
$\dot{M}_w$ also remain approximately unchanged when converting from numerical to
physical units ($\times0.3$ for M03 runs and $\times3$ for M3 runs). The results are consistent
with the Hall-free case studied in \citet{BaiStone13b}, indicating that the strength of the wind is
solely determined by the {\it physical strength} of the magnetic field. Also from Table \ref{tab:runs1},
accretion rate driven by magnetic braking is more or less unaffected by the variation of disk
surface density.

\subsubsection[]{Grain Abundance}

\begin{figure*}
    \centering
    \includegraphics[width=170mm]{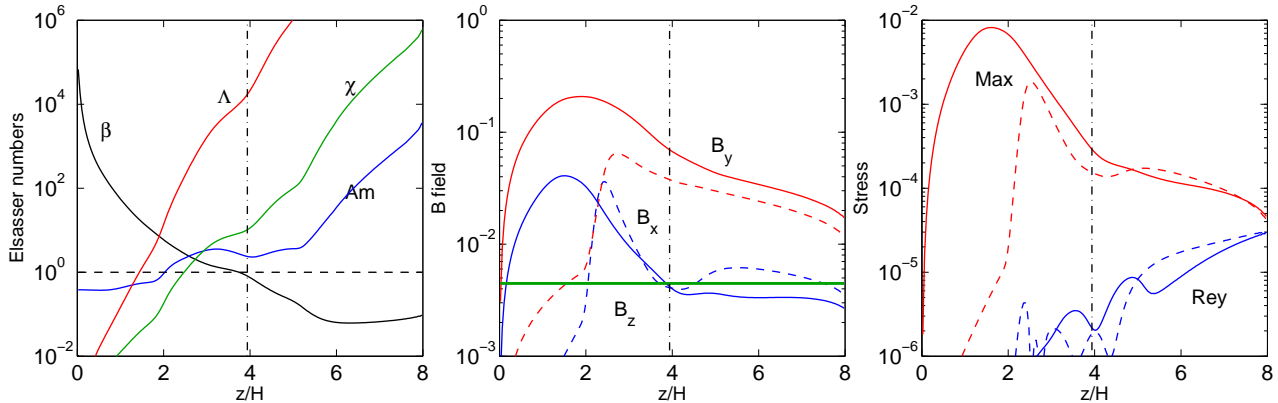}
  \caption{The vertical profiles of the Elsasser numbers (left), the three
  magnetic field components (middle) and the $R\phi$ components of Maxwell
  and Reynolds stress (right) from our grain-free run nogr-R1b5H+ (solid). In the
  middle and right panels, we also show in dashed lines the corresponding profiles
  from our fiducial run R1b5H+ (with grains) for comparison. The vertical dash-dotted
  line indicate the location of the wind base for run nogr-R1b5H+.}\label{fig:1AUpos}
\end{figure*}

We next consider a run using grain-free chemistry labeled by ``nogr". We find that the strength
of the disk wind, characterized by $T_{z\phi}^{\rm Max}$ and $\dot{M}_w$, is stronger than the
fiducial case for both $B_z>0$ and $B_z<0$ cases, although the enhancement is only modest,
which is again consistent with findings in \citet{BaiStone13b} for the Hall-free case. However, in
the case of $B_z>0$, the enhancement of $\alpha^{\rm Max}$ (magnetic braking) is substantial. In
Figure \ref{fig:1AUpos}, we show the corresponding Elsasser number, magnetic field and stress
profiles for run nogr-R1b5H+. We see that the midplane value of $Am$ increases by $\sim3$ orders
of magnitude compared with the fiducial run R1bb5H+, reflecting the increase in midplane ionization
fraction. The reduced magnetic diffusivity toward the disk midplane makes magnetic field amplification
discussed in Section \ref{ssec:relax} extend to much deeper regions than the fiducial case. We see
that it was not until very close to the midplane that the horizontal field starts to drop to zero by
enforced symmetry. The deeper penetration with continued amplification that acts to both $|B_x|$
and $|B_y|$, which leads to much stronger Maxwell stress, giving $\alpha^{\rm Max}$ about 10 times
larger than the fiducial case.

We can compare our grain-free simulation results with the results of \citet{Lesur_etal14}. They
conducted full-disk simulations with an analytical prescription of grain-free chemistry.
They obtained $\alpha^{\rm Max}\sim0.05$, compared with $\alpha^{\rm Max}\sim0.01$ in our case.
For fair comparison, we further performed a grain-free run with full disk, obtaining
$\alpha^{\rm max}\approx1.4\times10^{-2}$. This is very close to our half-disk simulation result
magnetic field amplification proceeds to the midplane in both cases. This value is a factor of
$\sim3$ smaller than their result mainly because our grain-free chemistry calculation is based on a 
complex chemical reaction network that yields smaller ionization fraction than their analytical formula
(checking the midplane $Am$ value indicates a factor of $\sim3$ difference).
\citet{Lesur_etal14} also reported that the midplane magnetic field is amplified to equipartition level
and strongly affects the disk hydrostatic equilibrium. In our full-disk grain-free simulation, we find the
total $\beta\sim10$ at disk midplane and drops below $1$ at $|z|\gtrsim2H$ (to affect hydrostatic
structure). This is again because of the lower ionization fraction from our grain-free chemistry.

\subsubsection[]{Ionization rate}

Finally, we vary the X-ray luminosity to $L_X=10^{29}$ and $10^{31}$ ergs s$^{-1}$ to study the
role of X-ray ionization on wind properties. The range of variation reflects the observed scatters of
X-ray luminosities in young stars \citep{Preibisch_etal05}, and may also account for the fact that
X-ray luminosities in young stars are highly variable \citep{Wolk_etal05}. We find that reducing the
X-ray luminosity only modifies the properties of the wind slightly, which is mainly because the wind
is launched from the surface layer dominated by FUV ionization. On the other hand, increasing
the X-ray luminosity leads to modest increase of the wind strength by allowing the wind to be launched
from deeper regions (due to enhanced ionization, see the values of $z_b$ in Table \ref{tab:runs1}).

\begin{figure}
    \centering
    \includegraphics[width=90mm]{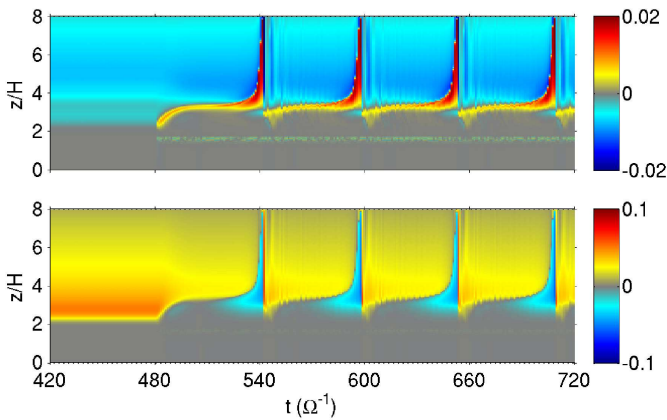}
  \caption{Time evolution of the magnetic field profile $B_x$ (upper panels) and $B_y$
  (lower panels) around the time the Hall effect is turned on at $t=480\Omega^{-1}$, for
  our run X31-R1b5H--, which shows periodic flips of horizontal fields.
  }\label{fig:X31neg}
\end{figure}

Most interestingly, we find that the wind solution becomes unstable in the $B_z<0$ case
with enhanced X-ray ionization\footnote{For $L_X=10^{29}$ with $B_z<0$, we find similar
unstable behavior for more subtle reasons.}. In Figure \ref{fig:X31neg} we show the time
evolution of the horizontal magnetic field in our run X31-R1b5H$-$. After turning on the
Hall term at $t=480\Omega^{-1}$, system quickly relaxes to a new configuration, but then
becomes unstable and the horizontal field flips in less than 10 orbits. This flip phenomenon
repeats itself quasi-periodically. The cause of the flip, as well as the consequences, will be
discussed in the next subsection in combination with a wider range of runs. Despite being
unstable, the bulk of the configuration still consists of a disk wind. The values shown in Table
\ref{tab:runs1} for this run are obtained by measuring the wind properties from $t=498-510$,
and we see from all major diagnostics that the strength of the wind is also modestly enhanced
compared with the Hall-free case.

\subsection[]{Parameter Space of Stable Wind Solutions}\label{ssec:stable}

We then perform a series of quasi-1D simulations with different $\beta_0$ and at different
disk radii. We find some of the quasi-1D runs never relax to a steady state. Instead, they
show similar behaviors as our run X31-R0b5$-$ with the large-scale horizontal field
changing sign quasi-periodically. Whenever this happens, steady state wind solution is
unlikely possible, and we do not include these runs in Table \ref{tab:runs2}.

\begin{figure}
    \centering
    \includegraphics[width=85mm]{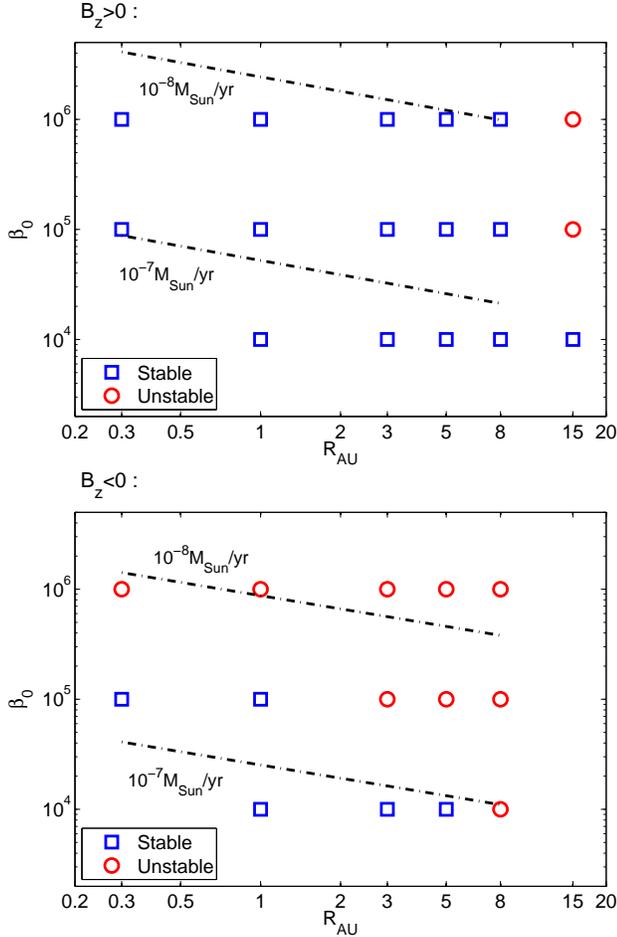}
  \caption{Parameter space in $R_{\rm AU}$ and $\beta_0$ (see Equation \ref{eq:beta0})
  where stable wind solutions can be found. Upper and lower panels are for $B_z>0$ and
  $B_z<0$ cases respectively. Stable regions are marked with blue squares. Regions where
  no stable wind solution can be found are marked by red circles. The two dashed lines
  indicate the desired $\beta_0$ as a function of radius for the wind-driven accretion rate
  to be $10^{-8}$ and $10^{-7}M_{\bigodot}$ yr$^{-1}$, based on the fitting formulas
  (\ref{eq:fit2}) and (\ref{eq:fit4}) in Section 5.4. 
  }\label{fig:stability}
\end{figure}

In Figure \ref{fig:stability} we show the stability map for all the quasi-1D simulations
we performed in the parameter exploration of $R_{\rm AU}$ and $\beta_0$.
Compared with the Hall-free situation of \citet{Bai13} (see his Figure 7), it is
clear that when $B_z>0$, the parameter space for stable wind solutions is
considerably enlarged: stable solutions can be found with weaker vertical field and
outer disk radii; while if $B_z<0$, the parameter space for stability is largely
reduced: stable solutions can only be found with stronger vertical field and at smaller
disk radii.

The onset of the instability in the unstable runs is due to the MRI, and the observed
stability trend can be readily understood from the Hall-MRI linear dispersion relation.

\begin{figure*}
    \centering
    \includegraphics[width=170mm]{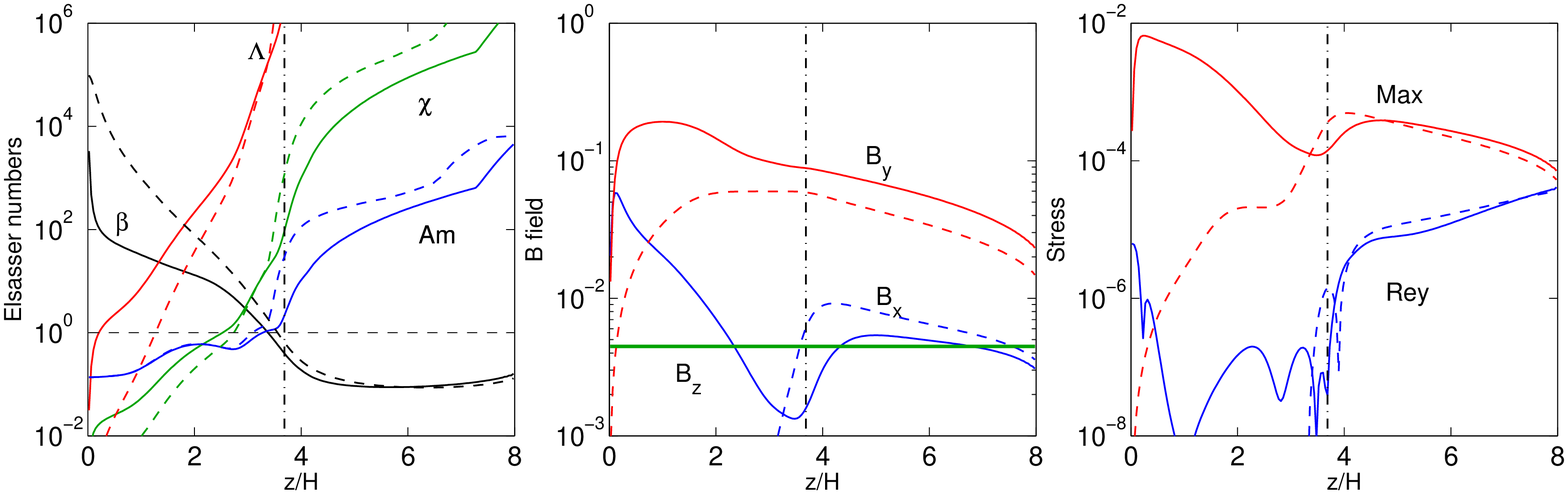}
  \caption{The vertical profiles of the Elsasser numbers (left), the three
  magnetic field components (middle) and the $R\phi$ components of Maxwell
  and Reynolds stress (right) from our runs R5b5H+ (solid) and R5b5H0 (dashed).
  The vertical dash-dotted line indicate the location of the wind base for run
  R5b5H+.}\label{fig:5AUpos}
\end{figure*}

As discussed in Section \ref{ssec:mriwind}, the main reason for the existence of stable
(Hall-free) wind solutions is that the MRI unstable modes under the given vertical field
strength become too long to fit into the disk. The main role played by the Hall effect is
that, for $B_z>0$, the unstable MRI modes shift to smaller $k_zv_{Az}$, or longer
wavelength at fixed vertical field (\citealp{Wardle99}, and see Figure \ref{fig:halldisp} in
the Appendix). Therefore, to make the system unstable, further weaker vertical field
(smaller $v_{Az}$) is required. This explains why the parameter space for stability is
enlarged when $B_z>0$.

For $B_z<0$ (or $\chi<0$), unstable MRI modes exist only when $|\chi|>1/2$.
Without dissipation, the unstable modes essentially extends to infinitely small scales
for $1/2<|\chi|<5/4$ \citep{Wardle99}. With dissipation, mostly ambipolar diffusion, the
unstable modes cutoff at finite wavelength, yet still extend to scales smaller than the
Hall-free case. This can again be seen from Figure \ref{fig:halldisp}
in the Appendix\footnote{Since the MRI dispersion relation in the presence of pure
vertical field is identical for the case with Ohmic resistivity and AD \citep{Wardle99},
one can simply replace $\Lambda$ by $Am$ in that figure to see the trend.}. Since
$|\chi|$ transitions from $\ll1$ to $\gg1$ from midplane to surface, it always falls in this
range at certain height. At this location, the unstable MRI modes
extend to shorter wavelength for fixed vertical field, making it more susceptible to the
MRI. Indeed, when checking with Figure
\ref{fig:X31neg} as well as many other unstable runs, we find that regions that first
lead to instability (at $z\sim3-3.5$ in Figure \ref{fig:X31neg}) are typically associated
with $|\chi|$ transition through order unity.

The above discussions are also consistent with the analysis of the MRI linear
modes presented by \citet{WardleSalmeron12}.

Connecting to the discussions at the end of the previous subsection (\ref{ssec:vari}),
we note that the range of stability can depend on other parameters such as grain
abundance and X-ray ionization. It is conceivable that, for example, with enhanced
X-ray ionization and reduced grain abundance, the range of radii where the laminar
wind solution holds would shrink, at least in the case of $B_z<0$. 

\subsection[]{Solutions toward Outer Radii}\label{ssec:outsol}

Toward outer disk radii, external ionization penetrates deeper into the disk
midplane. Together with reduced gas density, this leads to rapid increase of
the ionization fraction, making the disk midplane region better coupled to magnetic field.
Two consequences result. First, the wind is launched from a lower height compared
with our fiducial runs, leading to higher mass outflow rate and wind stress (in code units).
This has been discussed in \citet{Bai13}. Second, with the Hall effect, this makes
magnetic field amplification more prominent in the $B_z>0$ case.

In Figure \ref{fig:5AUpos}, we show the profiles of the Elsasser numbers, the
three magnetic field components, and the $R\phi$ stresses from our runs
R5b5H0 and R5b5H+ (note that run R5b5H$-$ is MRI unstable). We see that in the
Hall-free case, the radial field $B_x$ diminishes well before reaching the midplane. With
the Hall term, $B_x$ continues to increase toward the midplane until it catches up with
the toroidal field $B_y$ before diminishing to zero at the midplane enforced by symmetry.
The strongly enhanced $B_x$ combined with $B_y$ makes the Maxwell stress
$\alpha^{\rm Max}$ more than $30$ times larger than the Hall-free case. This example is
an exaggerated version of the 1AU fiducial runs discussed in detail in Section
\ref{ssec:relax}. It is similar to the grain-free run discussed in Section \ref{ssec:vari}, and 
also other wind solutions at comparable or larger radii listed in Table \ref{tab:runs2} are
qualitatively similar. We also note that at radius $R\geq5$ AU, the location of maximum
inflow velocity $z_{\rm in, max}$ is either very close or exactly at the disk midplane,
indicating that entire disk is actively coupled to the magnetic field.

\begin{figure*}
    \centering
    \subfigure{
    \includegraphics[width=150mm,height=60mm]{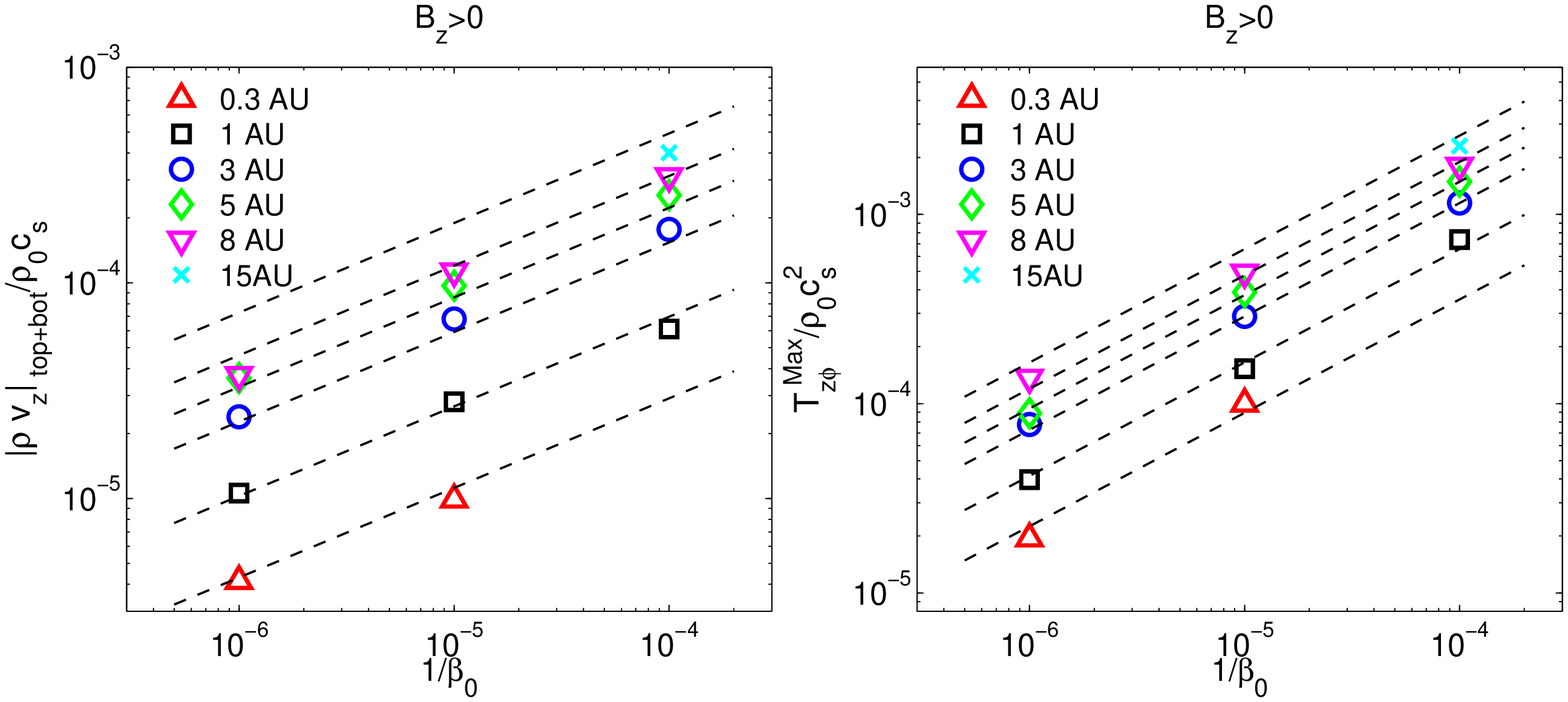}}
    \subfigure{
    \includegraphics[width=150mm,height=60mm]{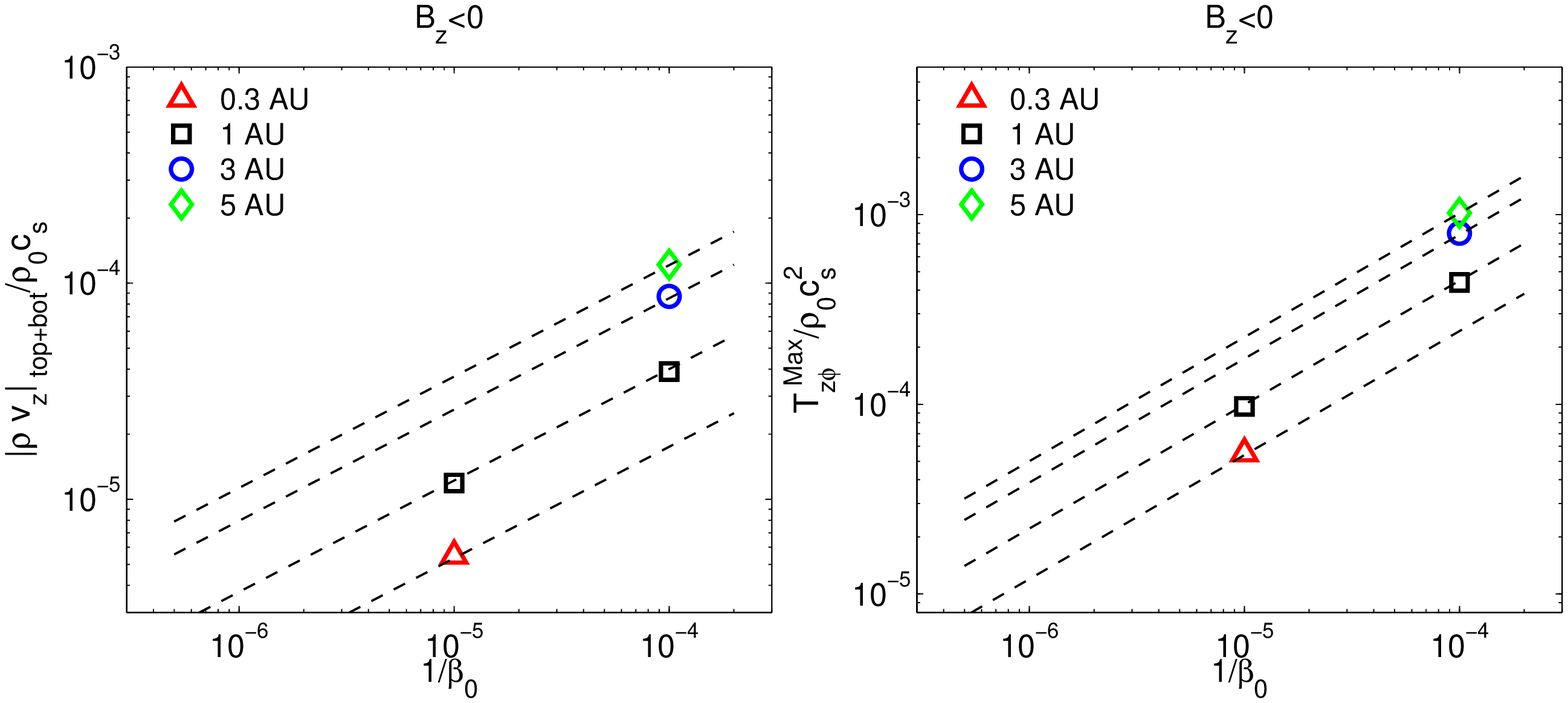}}
  \caption{Single-sided wind mass loss rate $\dot{M}_w$ (left) and the wind stress
  $T_{z\phi}^{\rm z_b}$ (right) from all simulations listed in Table \ref{tab:runs2}.
  Upper and lower panels correspond to simulations with $B_z>0$ and $B_z<0$ respectively.
  Also plotted in dashed lines are the fitting formulas (\ref{eq:fit1})-(\ref{eq:fit4}) applied
  to appropriate panels with individual lines corresponding to individual radius.}\label{fig:scaling}
\end{figure*}

Despite the stronger magnetic coupling in the midplane region, the properties of the
wind seem to be less affected, as we see that to the right of the vertical dash-dotted line
in Figure \ref{fig:5AUpos}, the magnetic profile, as well as the Maxwell and Reynolds
stresses behave in a way very similar to the 1 AU case.
 
\subsection[]{Angular Momentum Transport and Mass Outflow}\label{ssec:scale}

In this subsection, we follow the approach of \citet{Bai13} and study the dependence of
the wind-driven accretion rate on $R_{\rm AU}$ and $\beta_0$. Using all the data from
Table \ref{tab:runs2}, we fit the wind mass loss rate $\dot{M}_w$ and the wind stress
$T_{z\phi}^{z_b}$ in the form of $CR_{\rm AU}^{q}\beta_0^{-b}$, where $C$, $q$ and
$b$ are constants.

For $B_z>0$, there are a total of 15 data points, we find
\begin{equation}\label{eq:fit1}
\frac{\dot{M}_w}{\rho_0c_s}\approx2.67\times10^{-5}\bigg(\frac{\Sigma_{\rm MMSN}}{\Sigma}\bigg)
\bigg(\frac{R}{\rm AU}\bigg)^{0.72}\bigg(\frac{\beta_0}{10^5}\bigg)^{-0.42}\ ,
\end{equation}

\begin{equation}\label{eq:fit2}
\frac{T_{z\phi}^{z_b}}{\rho_0c_s^2}\approx1.65\times10^{-4}\bigg(\frac{\Sigma_{\rm MMSN}}{\Sigma}\bigg)
\bigg(\frac{R}{\rm AU}\bigg)^{0.51}\bigg(\frac{\beta_0}{10^5}\bigg)^{-0.60}\ .
\end{equation}

For $B_z<0$, although there are only 5 data points, we find a very tight fit
\begin{equation}\label{eq:fit3}
\frac{\dot{M}_w}{\rho_0c_s}\approx1.59\times10^{-5}\bigg(\frac{\Sigma_{\rm MMSN}}{\Sigma}\bigg)
\bigg(\frac{R}{\rm AU}\bigg)^{0.69}\bigg(\frac{\beta_0}{10^5}\bigg)^{-0.50}\ ,
\end{equation}

\begin{equation}\label{eq:fit4}
\frac{T_{z\phi}^{z_b}}{\rho_0c_s^2}\approx9.96\times10^{-5}\bigg(\frac{\Sigma_{\rm MMSN}}{\Sigma}\bigg)
\bigg(\frac{R}{\rm AU}\bigg)^{0.51}\bigg(\frac{\beta_0}{10^5}\bigg)^{-0.65}\ .
\end{equation}
In the above two formulas, uncertainties to all fitting coefficients are found to be less than
$5\%$. The inclusion of surface density in these relations is based on the discussions in
Section \ref{sssec:sigma}.

Without the Hall term, the fitting results are very close to Equations (9) and (10) of
\citet{Bai13}, thus we do not repeat. Also note that we have quoted single-sided mass
loss rate while \citet{Bai13} used mass loss rate from both sides.

In Figures \ref{fig:scaling}, we show the values of the wind mass loss rate and the
wind stress from all runs in Table \ref{tab:runs2} together with the above formulas.
We see that these formulas generally fit the data very well. The power law indices of
the scaling relations are similar indicating similar wind physics. We have also checked
the scalability of $\alpha^{\rm Max}$ and find that while it has the trend to
monotonically increase with increasing magnetic flux, the dependence on
$R_{\rm AU}$ is not monotonic.

Interpretation of these fitting formula follows from the discussions in Section 3.2 of
\citet{Bai13}. With proper unit conversion, i.e., Equation (\ref{eq:dotMV}), the wind
stress $T_{z\phi}^{z_b}$ should provide reliable estimates of the wind-driven
accretion rate. When applied to disks with surface density $\Sigma$, one should
interpret $\beta_0$ as the ratio of the midplane gas pressure of {\it a MMSN} disk
to the magnetic pressure of the net vertical field.
The wind mass loss rate $\dot{M}_w$, on the other hand, is not
well determined in the shearing-box framework
\citep{Fromang_etal13,BaiStone13b}. The normalization factors in the $\dot{M}_w$
fitting formulas are likely significant overestimated, while power-law indices are
probably more reliable, which at least sets the benchmark for shearing-box simulations.

Using these fitting formulas and Equation (\ref{eq:dotMV}) assuming MMSN disk,
we further show in Figure \ref{fig:stability} in dash-dotted lines the desired value
of $\beta_0$ as a function of $R_{\rm AU}$ for the disk to maintain wind-driven
accretion rate of $10^{-8}$ and $10^{-7}M_{\bigodot}$ yr$^{-1}$. We see that
given the typical accretion rate of $10^{-8}-10^{-7}M_{\bigodot}$ yr$^{-1}$, stable
wind solutions for disks with $B_z>0$ extend up to $\sim15$ AU, while for disks
with $B_z<0$, stable solutions exist only up to 3-5 AU. Reducing the accretion
rate would lead to reduced radial range of stability.

Based on the fact that wind-driven accretion rate depends only on the physical
strength of the net vertical field $B_z$, we can further write down the
wind-driven accretion rate in terms of the physical field strength
\begin{equation}
\dot{M}_V=0.82\times10^{-8}M_{\bigodot}\ {\rm yr}^{-1}R_{\rm AU}^{1.71}
\bigg(\frac{B_z}{10{\rm mG}}\bigg)^{1.2}\label{eq:MVpos}
\end{equation}
for $B_z>0$, and
\begin{equation}
\dot{M}_V=0.42\times10^{-8}M_{\bigodot}\ {\rm yr}^{-1}R_{\rm AU}^{1.87}
\bigg(\frac{B_z}{10{\rm mG}}\bigg)^{1.3}\label{eq:MVneg}
\end{equation}
for $B_z<0$.
These expressions can be considered as disk-model independent, as long as
a stable laminar-wind solution exists. The only MMSN scaling comes from the
temperature profile, which is reasonable for irradiated disks.

The above formulas are to be compared with the Hall-free formula based on
the results of \citet{Bai13} \footnote{We regret the miscalculation in Equation
(11) of \citet{Bai13}.}
\begin{equation}
\dot{M}_V=0.47\times10^{-8}M_{\bigodot}\ {\rm yr}^{-1}R_{\rm AU}^{1.90}
\bigg(\frac{B_z}{10{\rm mG}}\bigg)^{1.32}\ .
\end{equation}
These formulas again capsulate the role of the Hall term on the properties of the
laminar wind solutions. With similar scalings, they reveal the enhancement and 
reduction on the strength of disk wind introduced by the Hall term when
$B_z>0$ or $B_z<0$.

Our results indicate that $10-100$mG net vertical magnetic field is necessary to
achieve wind-driven accretion rate of $10^{-8}-10^{-7}M_{\bigodot}$ yr$^{-1}$ at
1 AU. Stronger field is required for the $B_z<0$ case and weaker field for
$B_z>0$. For steady-state accretion, the radial profile of $B_z$ should satisfy
$B_z\propto R^{-1.43}$ for both magnetic polarities, corresponding to magnetic
flux distribution of $\Phi(R)\propto R^{0.57}$, where $\Phi(R)$ denotes the total
magnetic flux contained within radius $R$.

\section[]{Summary and Discussion}\label{sec:conclusion}

\subsection[]{Summary}

In this work, we have successfully implemented the Hall term in the ATHENA
MHD code, which enables us to extend our previous study on the gas dynamics
of protoplanetary disks (PPDs) to include all three non-ideal MHD effects in a
self-consistent manner in local shearing-box simulations. All our simulations
include an external vertical magnetic field $B_z$, which has been realized to
be essential (see discussions in Section \ref{ssec:mriwind}). We focus on the inner
region of PPDs (up to $\sim10-15$ AU) in this paper where the disk is expected to
be largely laminar with accretion driven by a magnetocentrifugal wind.

Our first important finding is that including the Hall term, the conclusion
from our previous work (\citealp{BaiStone13b,Bai13} where only Ohmic resistivity
and ambipolar diffusion were included) that the inner disk is largely laminar still
holds, with accretion mainly driven by a magnetocentrifugal wind. On the other hand,
the wind solution is further controlled by the Hall effect in a way that depends on the
polarity of the external vertical field (sign of $B_z$), and we summarize as follows.

For external field being aligned with disk rotation ($B_z>0$), we find
\begin{itemize}
\item The horizontal magnetic field is strongly amplified compared with the
Hall-free solutions due to the Hall-shear instability, leading to stronger
disk wind and more efficient (vertical) angular momentum transport by up to
$\sim50\%$.

\item The enhanced horizontal magnetic field drives radial transport of angular
momentum via large-scale Maxwell stress (magnetic braking), which accounts
for a considerable fraction of the wind-driven accretion rate.

\item The parameter space where a stable laminar wind solution can be
found is extended. For typical accretion rates of $10^{-8}-10^{-7}M_{\bigodot}$
yr$^{-1}$, radial range of stability extends to $\sim10-15$ AU before the MRI sets in.
\end{itemize}

For external field being anti-aligned with disk rotation ($B_z<0$), we find
\begin{itemize}
\item The horizontal magnetic field is reduced compared with the Hall-free
solutions, leading to weaker disk wind and less efficient (vertical) angular
momentum transport by $\sim20\%$.

\item Radial transport of angular momentum by magnetic braking is negligible.

\item The parameter space for a stable laminar wind solution is substantially
reduced. For typical accretion rates of $10^{-8}-10^{-7}M_{\bigodot}$ yr$^{-1}$,
radial range of stability extends only to $\sim3-5$ AU before the MRI sets in.
\end{itemize}

For our fiducial simulation parameters (1AU with $\beta_0=10^5$), we explored
the dependence of the wind solutions on the disk surface density, X-ray ionization
rate and grain abundance. Our results indicate that the wind properties are largely
determined by the physical strength of the vertical magnetic field. They are
independent of the disk surface density and weakly dependent on the ionization
structure (e.g., grain abundance, ionization rate). On the other hand, the efficiency
of magnetic braking strongly depends on the diffusivity profiles, and grain-free
chemistry yields much higher $\alpha^{\rm Max}$ than calculations with grains.

Using the MMSN disk model, and using standard prescriptions of ionization
and chemistry, we further explored the dependence of the wind properties on
disk radii and the strength of the vertical magnetic field. The results are best
summarized in the fitting formulas (\ref{eq:fit1}) to (\ref{eq:fit4}). We also provide
disk model independent formulas for the wind-driven accretion rate in Equations
(\ref{eq:MVpos}) and (\ref{eq:MVneg}). Our results indicate that $10-100$mG
net vertical magnetic field at 1 AU is required to account for the typical PPD
accretion rates. At fixed accretion rate, stronger (by a factor of $\sim2$) net field
is needed in the $B_z<0$ case.

Most of our simulations cover half of the disk with enforced reflection boundary
condition at the disk midplane to guarantee that the wind solutions obey the
even-$z$ symmetry (horizontal magnetic field changes sign at midplane). We find
that in the $B_z<0$ case, these solutions can always be realized when using a full
disk thanks to the Hall effect, which reduces horizontal field strength toward the disk
midplane. When $B_z>0$, however, we are currently unable to achieve the even-$z$
wind solutions in full-disk simulations at $1$ AU. The system tends to end up at the
odd-$z$ symmetry solution where the outflow geometry is unphysical for a disk wind.
This solution gives stronger magnetic braking by a factor of up to a few since
horizontal magnetic field is amplified throughout the midplane instead of passing
through zero. While it is unclear which solution the nature picks due to limitations
of the shearing-box framework, we do find that the even-$z$ symmetry wind solution
can be realized at slightly larger disk radii ($\sim3-5$ AU), which we will discuss
in our companion paper. Finally, the even-$z$ wind solutions obtained via
simulations tend to have large radial drift velocities of magnetic flux, which is
likely affected by vertical boundary conditions. Global simulations are essential to
resolve the symmetry issues and to yield realistic rate of magnetic flux transport.

\subsection[]{Discussions}

Our work strengthens the notion that the evolution of PPDs is largely governed
by the distribution and transport of external poloidal magnetic flux, as suggested
in our previous works \citep{BaiStone13b,Bai13}. Such large-scale field is
expected as a natural consequence of star formation: molecular clouds and
star-forming cores are all strongly magnetized (e.g., see \citealp{Crutcher12} for
a review). Recent dust polarization observations further reveal the presence of
large-scale field threading protostellar cores. The large-scale fields appear to be
randomly oriented with respect to the direction of protostellar outflows
\citep{Hull_etal14}, while there is also evidence of preferential alignment for
more isolated sources with projection effects taken into account
\citep{Chapman_etal13}. Considering the Hall effect, the bifurcation of disk wind
properties with different polarities of the external magnetic field further suggests
that PPDs may evolve differently with different initial magnetic field polarities, and
systems with ${\mb B}\cdot{\mb\Omega}>0$ may achieve higher accretion rate,
or retain less magnetic flux (in a self-organized way to avoid accreting to fast)
compared with systems with ${\mb B}\cdot{\mb\Omega}<0$.

On the other hand, the desired level of magnetic flux threading PPDs (at least in
the Class II phase) is tiny compared with the amount of magnetic
flux threading star-forming cores. Consequently, the process of star formation must
also be accompanied by the removal of magnetic flux, which is another major
problem in the theory of star formation. It appears that substantial magnetic flux must
be removed in order to form the PPD itself to avoid the ``magnetic braking
catastrophe" \citep{MellonLi08}, which may be achieved via misaligned magnetic
field \citep{HennebelleCiardi09,Joos_etal12}, external turbulence
\citep{SantosLima_etal12,Seifried_etal12} or non-ideal MHD effects
\citep{Li_etal11,Krasnopolsky_etal11,Tomida_etal13}. After disk formation, the
transport of magnetic flux must be achieved within the disk itself, yet the problem is
intrinsically global, and must depend on the overall magnetic field geometry and
internal dissipation in the disk (e.g. \citealp{Lubow_etal94}). Recently, there have been
several semi-analytical works to study magnetic flux transport in thin accretion
disks which have revealed complex dependence on the internal disk microphysics
(e.g., \citealp{GuiletOgilvie12,GuiletOgilvie13,Okuzumi_etal13,TakeuchiOkuzumi13}).
Nevertheless, a still missing important ingredient is the launching of magnetic outflow,
and it appears that global simulations of PPDs with resolved disk microphysics and
sufficiently large vertical domain to accommodate disk outflow is essential toward
a better understanding.

While our results suggest the inner region of PPDs is largely laminar in terms of
magnetic activities, it does not exclude the possibility for pure hydrodynamic
mechanisms to generate turbulence. In fact, some level of turbulence is probably
needed to keep at least some small dust suspended in the disk so as to explain
the scattered starlight as well as the near infrared spectral energy distribution (e.g.,
\citealp{Stapelfeldt_etal03,DAlessio_etal06}). Promising candidates of hydrodynamic
turbulence may include the Goldreich-Schubert-Fricke instability
\citep{Urpin03,Nelson_etal13}, the critical layer instability \citep{Marcus_etal13} and
the convective overstability and baroclinic vortex amplification
\citep{Petersen_etal07b,LesurPapaloizou10,Raettig_etal13,KlahrHubbard14,Lyra14}.
We note that the survival of these instabilities requires the suppression of MRI (e.g.,
\citealp{LyraKlahr11}), hence the inner disk is a very promising location for them
to operate, provided that appropriate thermodynamic conditions are met.

A largely laminar inner disk is favorable for many processes of planet formation,
including grain growth, planetesimal formation and further growth toward planetary
embryos, which were discussed in \citet{BaiStone13b}. A largely-laminar disk may
further alleviate the problem of type-I migration by allowing low-mass planets/cores to
open gaps \citep{GoodmanRafikov01,DongRafikov11b}, which may substantially
reduce the rate of inward migration. The polarity dependence of wind properties and
the MRI stability threshold might also indicate that planet formation and retention are
dependent on the polarity of the initial large-scale field, although the details need to
be filled up.

While our simulations have captured the most essential non-ideal MHD physics in the
inner region of PPDs, there are still several unresolved problems due to limitations of
the shearing-box framework. Important issues include the large-scale kinematics of
the wind launched from the disk, wind geometry and symmetry, and the direction of
magnetic flux transport. The next step forward would be to address these issues with
global simulations, which are planned as our future work. In the companion paper, we
again take the advantage of the shearing-box for its resolving power of disk
microphysics and study the Hall-controlled gas dynamics in the outer PPDs.

\acknowledgments

I thank the referee for a very thoughtful report with helpful suggestions that greatly
improve the presentation of this paper. I am also grateful to Jim Stone and Arieh Konigl
for helpful discussions, to Ruth Murray-Clay for useful conversations and a proof-reading.
I thank H. Nomura for helpful discussions on the FUV ionization/chemistry at disk surface
and together with C. Walsh for providing their calculation data. This work is supported for
program number HST-HF-51301.01-A provided by NASA through a Hubble Fellowship
grant from the Space Telescope Science Institute awarded to XN.B, which is operated by
the Association of Universities for Research in Astronomy, Incorporated, under NASA
contract NAS5-26555.

\appendix

\section[]{Implementation of the Hall Term in the ATHENA MHD code}

The non-dissipative nature of the Hall term makes its implementation not as
straightforward as Ohmic resistivity and ambipolar diffusion. It is well known that
first and second order explicit schemes are unconditionally unstable
\citep{Falle03,KunzLesur13}. Here we consider the alternative scheme suggested
by \citet{OSullivanDownes06,OSullivanDownes07}, who showed that dimensionally
split method makes a stable Hall-MHD algorithm.

For simplicity, we consider the one-dimensional algorithm. We assume the background
magnetic field $B_0$ is along the $x$-axis. Due to the Hall term, the magnetic fields
evolve as
\begin{equation}
\begin{split}
\frac{\pa B_y}{\pa t}&=Q_HB_0\frac{\pa^2}{\pa x^2}B_z\ ,\\
\frac{\pa B_z}{\pa t}&=-Q_HB_0\frac{\pa^2}{\pa x^2}B_y\ ,
\end{split}
\end{equation}
where $Q_H=\eta_H/B_0$ is the coefficient for the Hall term. To update the magnetic
field from step $(n)$ to $(n+1)$, we first update $B_y^{(n)}$ to $B_y^{(n+1)}$ using
$B_z^{(n)}$, and then update $B_z^{(n)}$ to $B_z^{(n+1)}$ using $B_y^{(n+1)}$. To
show that this method is numerically stable, we perform von-Neumann analysis and
decompose magnetic perturbations into Fourier modes. Picking up an arbitrary mode,
and assuming $B_y=A_2\exp({\rm i}\omega t)\exp(-{\rm i}kx)$ and
$B_z=A_3\exp({\rm i}\omega t)\exp(-{\rm i}kx)$, we obtain
\begin{equation}
\begin{split}
A_2[\exp({\rm i}\omega\Delta t)-1]&=A_3D_H[2\cos(k\Delta x)-2]\ ,\\
A_3[\exp({\rm i}\omega\Delta t)-1]&=-A_2\exp({\rm i}\omega\Delta t)D_H[2\cos(k\Delta x)-2]\ ,\
\end{split}
\end{equation}
where $\Delta x$ and $\Delta t$ represent grid spacing and timestep, and
$D_H\equiv Q_HB_0\Delta t/\Delta x^2$. Non-trivial solutions demand
$A_2=\pm{\rm i}A_3\exp(-{\rm i}\omega\Delta t/2)$, and
\begin{equation}
\sin\frac{\omega\Delta t}{2}=\pm2D_H\sin^2\frac{k\Delta x}{2}\ .
\end{equation}
We see that for any given $k$, $\omega$ is a real number provided that $2|D_H|\leq1$,
hence the amplitude of the wave is preserved without damping or amplification. The
stability constraint is thus given by $|D_H|<1/2$, or
\begin{equation}
\Delta t_{\rm Hall,0}\leq\frac{\Delta x^2}{2|\eta_H|}\ .\label{eq:cfl_hall}
\end{equation}

Using this method, the Hall MHD term is implemented to ATHENA in a operator-split
manner. Since ATHENA uses the standard constrained transport (CT) to preserve the
divergence free condition, the actual procedure in our implementation follows the same
spirit of dimensionally-split update, with the split acting on the Hall electric field
${\mb E}^H=Q_H{\mb J}\times{\mb B}$. We first calculate $E^H_x$ using the original
magnetic field. Using $E^H_x$ alone, we update $B_y$ and $B_z$ for a full timestep,
from which we calculate $E^{H}_y$ using the original $B_x$ and updated $B_y$ and
$B_z$. Using $E^H_y$ alone, we further update $B_x$ and $B_z$ for a full timestep.
Finally, using the updated field components, we evaluate $E^{H}_z$. The obtained
Hall electric fields are then combined with the electric fields from Ohmic and AD terms 
to update the magnetic fields via CT.

With shearing-box, we also remap of $J_y$ at radial (shearing-box) boundaries so that
the line integral of $J_y$ along the azimuthal direction in the inner and outer radial
boundaries are equal. This is necessary to avoid numerical instabilities at radial
boundaries, which we have found earlier in the case with the ambipolar diffusion term
\citep{Simon_etal13a}, as well as the Hall term in the context of plasma simulations
\citep{Kunz_etal14}.

In multi-dimensions, the stability criterion becomes more stringent, and also depends
on the details of the implementation, which is particularly complicated by the CT
algorithm required in the ATHENA MHD code. Using the test problems described in the
next Appendix, the following stability criterion is found to be robust
\begin{equation}
\Delta t_{\rm Hall,1}\leq\frac{\Delta x^2}{2d|\eta_H|}\ ,\label{eq:cfl_hall2}
\end{equation}
where $d=1, 2, 3$ represents the dimension of the problem. This is the analog of the
stability criterion of a pure diffusion problem (e.g., Ohmic resistivity $\eta_O$), where the
timestep constraint is $\Delta t\leq\Delta x^2/4d\eta_O$.

Finally, we discuss the timestepping in the presence of all three non-ideal MHD terms,
which are all implemented in an operator-split manner.
The Ohmic and AD terms are parabolic in nature and can be treated jointly as magnetic
diffusion with total diffusivity $\eta_{\rm tot}=\eta_O+\eta_A$, which gives the diffusion
timestep constraint $\Delta t_{\rm diff}<\Delta x^2/4d\eta_{\rm tot}$. The small timestep
constraint can be relaxed by applying the super-timestepping (STS) technique
\citep{Alexiades_etal96}, where one employs multiple sub-steps of decreasing length
within a super timestep. The initial length of the sub-steps can be significantly larger than
the stability constraint $\Delta t_{\rm diff}$, but it is later stabilized by progressively small
sub-step lengths. This technique has been shown to be very successful in accelerating
the calculations with AD \citep{OSullivanDownes06,OSullivanDownes07,Choi_etal09},
and has been implemented and effectively used in our previous works
\citep{Bai12,Simon_etal13a,BaiStone13b,Bai13,Simon_etal13b}.

Due to the hyperbolic nature of the Hall term, the STS technique can not be used to
accelerate the calculation. However, we can still use STS to accelerate the calculation
for Ohmic resistivity and AD terms. The overall MHD timestep $\Delta t_{\rm all}$ is
determined by the minimum of the normal MHD timestep $\Delta t_{\rm MHD}$ (given by
the Courant-Friedrichs-Lewy condition) and $\Delta t_{\rm Hall,1}$ (\ref{eq:cfl_hall2}),
and the diffusion timestep is given by $\Delta t_{\rm diff}$. However, we find that the
in the presence of strong diffusion, the Hall timestep can be relaxed towards the 1D
criterion $\Delta t_{\rm Hall,0}$ (\ref{eq:cfl_hall}). Empirically, we adopt the full MHD
timestep to be
\begin{equation}
\Delta t_{\rm all}={\rm MIN}\bigg\{\Delta t_{\rm MHD}, {\rm MIN}\bigg[\Delta t_{\rm Hall,1}+
(\Delta t_{\rm Hall,0}-\Delta t_{\rm Hall,1})\frac{\eta_{\rm tot}}{\eta_H},
\Delta t_{\rm Hall,0}\bigg]\bigg\}\ .
\end{equation}
For Ohmic and AD terms, we use STS when $\Delta t_{\rm diff}<\Delta t_{\rm all}$, with
details the same as described in Appendix B.3.1 of Bai (2012, PhD thesis), repeated in
Appendix A of our later publication \citep{Simon_etal13a}.

We note that a different Hall algorithm was implemented by \citet{Lesur_etal14} following
\citet{Toth_etal08} using a whistler modified HLL Riemann solver. This Godunov
approach makes the Hall-MHD algorithm very robust, although the HLL solver itself is
very diffusive. Our operator-split algorithm is more flexible and is combined with the more
accurate and much less diffusive HLLD solver. On the other hand, being a marginally
stable algorithm, some level of external dissipation is generally needed if the system
becomes non-linear. Since the Hall effect is always accompanied by strong Ohmic
resistivity and/or AD in PPDs, our method is well suited for studying the gas dynamics in PPDs.

\section[]{Code Tests}

We describe two sensitive test problems to demonstrate the successful implementation
of the Hall term in ATHENA, where the first problem is reproduced from Appendix B.4.3
of Bai (2012, PhD thesis).

\subsection[]{Circularly Polarized Aflv\'en Wave Test}

In the presence of the Hall effect, left and right polarized Alfv\'en waves propagate
at different velocities, which makes it an excellent code test problem. Consider a
uniform medium with density $\rho_0$ and electron density $n_e$, embedded in a
uniform magnetic field ${\mb B}_0$. For Alfv\'en mode propagating along
${\mb B}_0$, the dispersion relation reads
\begin{equation}
\omega^2-k^2v_{Az}^2=\pm\omega k^2\frac{cB_{0}}
{4\pi n_ee}\ ,\label{eq:whistlerDR}
\end{equation}
where the plus (minus) sign corresponds to right (left) hand polarizations. The above
dispersion relation can be rewritten into a more intuitive form as
\begin{equation}
\omega^2=\bigg(1\pm\frac{\omega}{\omega_H}\bigg)k^2v_{A}^2\ ,\label{eq:whistlerDR2}
\end{equation}
where $\omega_H$ is the Hall frequency defined in Equation (\ref{eq:omgH}), and it has the
clear meaning of being the cut-off frequency for left-handed waves. The right
handed wave is also known as whistler wave and has the asymptotic
dispersion property of $\omega\propto k^2$ (for $\omega\gg\omega_H$).
Normalizing the wave number by $x\equiv kv_A/\omega_H=kl_H$, the phase
velocity is given by
\begin{equation}
\frac{v_{\rm ph}}{v_A}=\frac{\sqrt{x^2+4}\pm x}{2}\ ,\label{eq:vphix}
\end{equation}
where again the plus/minus sign corresponds to right/left handed Alfv\'en waves.

\begin{figure}
    \centering
    \includegraphics[width=160mm]{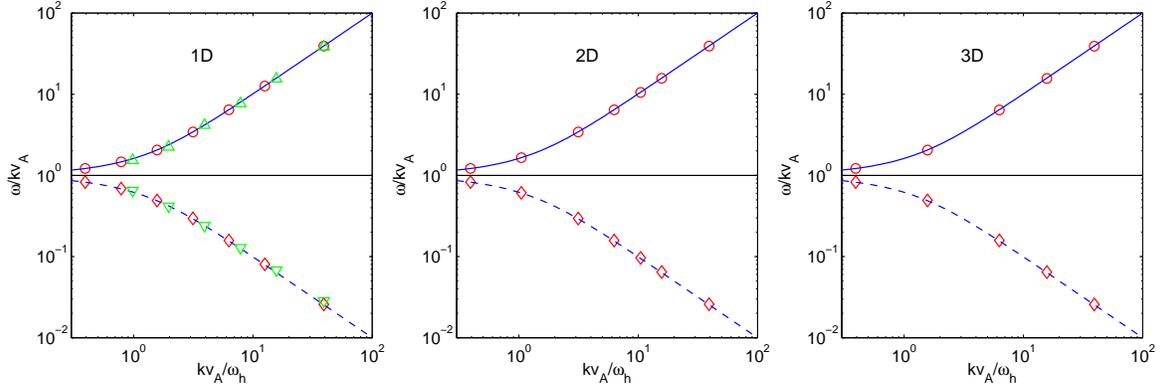}
  \caption{The measured dispersion relation for circularly polarized
  Alfv\'ven waves in 1D (left), 2D (middle) and 3D (right) grids. Upper
  panels show the results for right handed (whistler) waves, where
  red circles mark the measured phase velocity $v_{ph}=\omega/k$
  normalized by $v_A$ at various wave numbers $k$ (normalized by
  $\omega_H/v_A$), and solid blue line indicated the theoretical
  relation. Lower panels are for left handed waves, with red diamonds
  and blue dashed lines mark the measured and theoretical dispersion
  relations respectively. The green symbols in the 1D plot show test
  results using a reduced resolution of 12 cells per wavelength.}\label{fig:cpaw}
\end{figure}

We test the linear dispersion relation (\ref{eq:vphix}) by initializing the
exact wave eigenvector in a periodic box in 1D, 2D and 3D, with wave
amplitude $|\delta{\mb B}|=10^{-4}|{\mb B}|$. In 1D, the wave is
grid-aligned with wavelength of $1$ resolved by $32$ cells. In 2D and
3D tests, the wave vectors are not aligned with the grid, and we properly
choose box sizes so that the wavelength is also 1: In 2D, the box size is
$(\sqrt{5}, \sqrt{5}/2)$ resolved by $64\times32$ cells and in 3D, the box
size is $(3, 1.5, 1.5)$ resolved by $64\times32\times32$ cells. In Figure
\ref{fig:cpaw}, we show the measured dispersion relation for right (whistler)
and left handed Alfv\'en waves and compare them with analytical relations.
We see that the agreement is excellent in all cases. In particular, we are
able to resolve the whistler wave branch up to very large $k$. Benefited
from the low level of dissipation, the code can also well reproduce the
dispersion relation for both waves at much lower resolution, 12 cells per
wavelength, as shown in an additional 1D test with green symbols.

\subsection[]{Linear Growth Rate of the Magnetorotational Instability}

The second problem aims at testing the coupling between the Hall term and
rotation/shear in the context of shearing-box simulations. The test problem
is adopted from \citet{SanoStone02a}, where we compare the numerical
dispersion relation of the MRI with predictions from linear theory. We set up
a 3D unstratified shearing-box (cf. Section 2.2 but ignore vertical gravity)
threaded by a weak net vertical magnetic field $B_0$ corresponding to
plasma $\beta_0=800$. For this test problem, we include both Ohmic
resistivity and the Hall term, and consider axisymmetric perturbations of the
form $\propto\exp{({\rm i}kz+\sigma t)}$. The linear dispersion then reads
\citep{Wardle99,BalbusTerquem01}
\begin{equation}
\sigma^4+\frac{2k^2}{\Lambda}\sigma^3+\mathscr{E}_2\sigma^2
+\frac{2k^2}{\Lambda}(k^2+1)\sigma+\mathscr{E}_0=0\ ,\label{eq:halldisp}
\end{equation}
where
\begin{equation}
\mathscr{E}_2=2k^{2}+1+\frac{k^2}{\Lambda^2}
+\frac{k^{2}}{2\chi}
\bigg(\frac{2k^{2}}{\chi}-3\bigg)\ ,
\end{equation}
\begin{equation}
\mathscr{E}_0=
\frac{k^4}{\Lambda^2}
+k^{2}\bigg(k^{2}+\frac{2k^{2}}{\chi}-3\bigg)
\bigg(1+\frac{1}{2\chi}\bigg)\ .
\end{equation}
Here $\Lambda$ and $\chi$ are the Ohmic and Hall Elsasser numbers defined in
(\ref{eq:Elsasser}), based on the background net vertical field $B_0$, and $k$ is
normalized to $\Omega/v_A$.
The Hall Elsasser number $\chi$ can be positive or negative when the vertical
field $B_0$ is parallel or anti-parallel to rotation axis, while $\Lambda$ is always
positive. For pure Hall MRI, the linear dispersion relation above has the
property that unstable mode exists only when $1/\chi>-2$ \citep{Wardle99}, which
we will test, though this is not the case for more general perturbations
\citep{BalbusTerquem01}.

\begin{figure*}
    \centering
    \includegraphics[width=160mm]{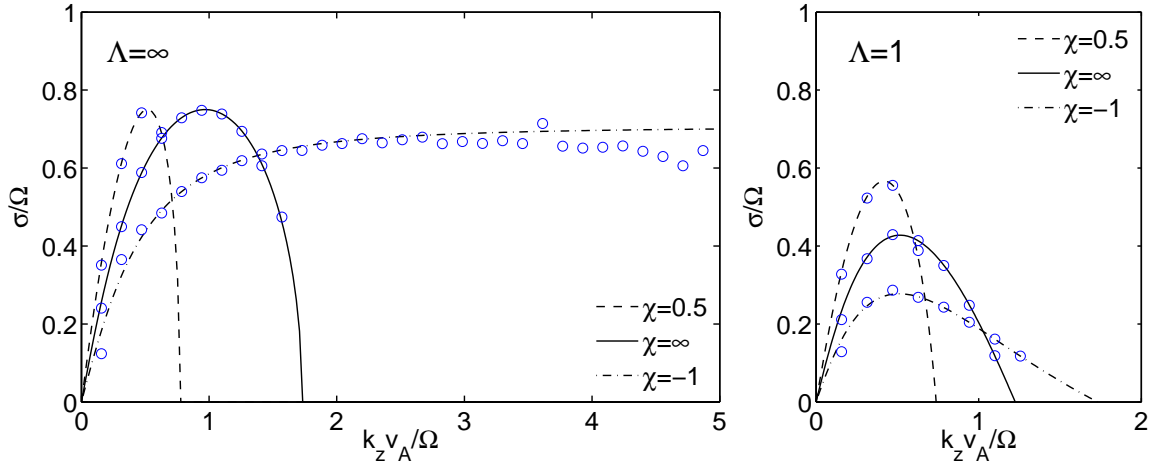}
  \caption{Linear dispersion relation of the MRI growth rate with Ohmic resistivity
  and the Hall effect. The growth rate $\sigma$ is normalized to the orbital frequency
  $\Omega^{-1}$, and is plotted as a function of $k_zv_A/\Omega$.
  }\label{fig:halldisp}
\end{figure*}

We adopt a very slim simulation box $L_x\times L_y\times L_z$=$0.1H\times0.1H\times2H$
resolved by $4\times4\times256$ cells, since we are interested in the vertical wave numbers.
Following \citet{SanoStone02a}, we initialize the problem with uniform gas density $\rho=1$
and random velocity perturbations (white noise) on the order of $\delta v=10^{-6}c_s$. We run
the test simulations with given Ohmic and Hall Elsasser numbers as input parameters for
about 2.5 orbits. From the simulations we perform Fourier analysis and evaluate the growth
rate of $v_x$ for every single vertical mode $k_z$ from time $60\Omega^{-1}$ to
$75\Omega^{-1}$. In Figure \ref{fig:halldisp}, we show the analytical growth rate versus
numerical growth rate from our test simulations. Note that the parameter $X$ adopted in
\citet{SanoStone02a} corresponds to $2/\chi$ in our case. We have considered parameters
with $\chi=0.5, \infty$ and $-1$, corresponding to $X=4, 0$ and $-2$ in \citet{SanoStone02a}.
We see that the numerical dispersion relation agrees very well with
analytical results. Particularly, for $\chi=-1$, MRI growth rate is well reproduced toward very
small wavelength due to the low level of intrinsic dissipation. This test demonstrates that our
implementation of the Hall term is well suited for conducting simulations in the shearing-box
framework.

\bibliographystyle{apj}

\begin{thebibliography}{125}
\expandafter\ifx\csname natexlab\endcsname\relax\def\natexlab#1{#1}\fi

\bibitem[{{Alexiades} {et~al.}(1996){Alexiades}, {Amiez}, \&
  {Gremaud}}]{Alexiades_etal96}
{Alexiades}, V., {Amiez}, G., \& {Gremaud}, P. 1996, Communications in
  Numerical Methods in Engineering, 12, 31

\bibitem[{{Bai}(2011{\natexlab{a}})}]{Bai11a}
{Bai}, X.-N. 2011{\natexlab{a}}, \apj, 739, 50

\bibitem[{{Bai}(2011{\natexlab{b}})}]{Bai11b}
---. 2011{\natexlab{b}}, \apj, 739, 51

\bibitem[{{Bai}(2012)}]{Bai12}
---. 2012, PhD thesis, Princeton University

\bibitem[{{Bai}(2013)}]{Bai13}
---. 2013, \apj, 772, 96

\bibitem[{{Bai} \& {Goodman}(2009)}]{BaiGoodman09}
{Bai}, X.-N. \& {Goodman}, J. 2009, \apj, 701, 737

\bibitem[{{Bai} \& {Stone}(2010{\natexlab{a}})}]{BaiStone10b}
{Bai}, X.-N. \& {Stone}, J.~M. 2010{\natexlab{a}}, \apj, 722, 1437

\bibitem[{{Bai} \& {Stone}(2010{\natexlab{b}})}]{BaiStone10c}
---. 2010{\natexlab{b}}, \apjl, 722, L220

\bibitem[{{Bai} \& {Stone}(2011)}]{BaiStone11}
---. 2011, \apj, 736, 144

\bibitem[{{Bai} \& {Stone}(2013{\natexlab{a}})}]{BaiStone13a}
---. 2013{\natexlab{a}}, \apj, 767, 30

\bibitem[{{Bai} \& {Stone}(2013{\natexlab{b}})}]{BaiStone13b}
---. 2013{\natexlab{b}}, \apj, 769, 76

\bibitem[{{Balbus} \& {Hawley}(1991)}]{BH91}
{Balbus}, S.~A. \& {Hawley}, J.~F. 1991, \apj, 376, 214

\bibitem[{{Balbus} \& {Terquem}(2001)}]{BalbusTerquem01}
{Balbus}, S.~A. \& {Terquem}, C. 2001, \apj, 552, 235

\bibitem[{{Birnstiel} {et~al.}(2012){Birnstiel}, {Klahr}, \&
  {Ercolano}}]{Birnstiel_etal12}
{Birnstiel}, T., {Klahr}, H., \& {Ercolano}, B. 2012, \aap, 539, A148

\bibitem[{{Blaes} \& {Balbus}(1994)}]{BlaesBalbus94}
{Blaes}, O.~M. \& {Balbus}, S.~A. 1994, \apj, 421, 163

\bibitem[{{Blandford} \& {Payne}(1982)}]{BlandfordPayne82}
{Blandford}, R.~D. \& {Payne}, D.~G. 1982, \mnras, 199, 883

\bibitem[{{Chapman} {et~al.}(2013){Chapman}, {Davidson}, {Goldsmith}, {Houde},
  {Kwon}, {Li}, {Looney}, {Matthews}, {Matthews}, {Novak}, {Peng},
  {Vaillancourt}, \& {Volgenau}}]{Chapman_etal13}
{Chapman}, N.~L., {Davidson}, J.~A., {Goldsmith}, P.~F., {Houde}, M., {Kwon},
  W., {Li}, Z.-Y., {Looney}, L.~W., {Matthews}, B., {Matthews}, T.~G., {Novak},
  G., {Peng}, R., {Vaillancourt}, J.~E., \& {Volgenau}, N.~H. 2013, \apj, 770,
  151

\bibitem[{{Choi} {et~al.}(2009){Choi}, {Kim}, \& {Wiita}}]{Choi_etal09}
{Choi}, E., {Kim}, J., \& {Wiita}, P.~J. 2009, \apjs, 181, 413

\bibitem[{{Crutcher}(2012)}]{Crutcher12}
{Crutcher}, R.~M. 2012, \araa, 50, 29

\bibitem[{{D'Alessio} {et~al.}(2006){D'Alessio}, {Calvet}, {Hartmann},
  {Franco-Hern{\'a}ndez}, \& {Serv{\'{\i}}n}}]{DAlessio_etal06}
{D'Alessio}, P., {Calvet}, N., {Hartmann}, L., {Franco-Hern{\'a}ndez}, R., \&
  {Serv{\'{\i}}n}, H. 2006, \apj, 638, 314

\bibitem[{{Davis} {et~al.}(2010){Davis}, {Stone}, \& {Pessah}}]{Davis_etal10}
{Davis}, S.~W., {Stone}, J.~M., \& {Pessah}, M.~E. 2010, \apj, 713, 52

\bibitem[{{Desch}(2004)}]{Desch04}
{Desch}, S.~J. 2004, \apj, 608, 509

\bibitem[{{Dong} {et~al.}(2011){Dong}, {Rafikov}, \& {Stone}}]{DongRafikov11b}
{Dong}, R., {Rafikov}, R.~R., \& {Stone}, J.~M. 2011, \apj, 741, 57

\bibitem[{{Ercolano} \& {Glassgold}(2013)}]{ErcolanoGlassgold13}
{Ercolano}, B. \& {Glassgold}, A.~E. 2013, \mnras, 436, 3446

\bibitem[{{Falle}(2003)}]{Falle03}
{Falle}, S.~A.~E.~G. 2003, \mnras, 344, 1210

\bibitem[{{Ferreira} \& {Pelletier}(1995)}]{FerreiraPelletier95}
{Ferreira}, J. \& {Pelletier}, G. 1995, \aap, 295, 807

\bibitem[{{Fleming} \& {Stone}(2003)}]{FlemingStone03}
{Fleming}, T. \& {Stone}, J.~M. 2003, \apj, 585, 908

\bibitem[{{Fleming} {et~al.}(2000){Fleming}, {Stone}, \&
  {Hawley}}]{Fleming_etal00}
{Fleming}, T.~P., {Stone}, J.~M., \& {Hawley}, J.~F. 2000, \apj, 530, 464

\bibitem[{{Fromang} {et~al.}(2013){Fromang}, {Latter}, {Lesur}, \&
  {Ogilvie}}]{Fromang_etal13}
{Fromang}, S., {Latter}, H., {Lesur}, G., \& {Ogilvie}, G.~I. 2013, \aap, 552,
  A71

\bibitem[{{Fromang} {et~al.}(2002){Fromang}, {Terquem}, \&
  {Balbus}}]{Fromang_etal02}
{Fromang}, S., {Terquem}, C., \& {Balbus}, S.~A. 2002, \mnras, 329, 18

\bibitem[{{Gammie}(1996)}]{Gammie96}
{Gammie}, C.~F. 1996, \apj, 457, 355

\bibitem[{{Garaud}(2007)}]{Garaud07}
{Garaud}, P. 2007, \apj, 671, 2091

\bibitem[{{Gardiner} \& {Stone}(2005)}]{GardinerStone05}
{Gardiner}, T.~A. \& {Stone}, J.~M. 2005, Journal of Computational Physics,
  205, 509

\bibitem[{{Gardiner} \& {Stone}(2008)}]{GardinerStone08}
---. 2008, Journal of Computational Physics, 227, 4123

\bibitem[{{Goldreich} \& {Lynden-Bell}(1965)}]{GoldreichLyndenBell65}
{Goldreich}, P. \& {Lynden-Bell}, D. 1965, \mnras, 130, 125

\bibitem[{{Goodman} \& {Rafikov}(2001)}]{GoodmanRafikov01}
{Goodman}, J. \& {Rafikov}, R.~R. 2001, \apj, 552, 793

\bibitem[{{Gressel} {et~al.}(2013){Gressel}, {Nelson}, {Turner}, \&
  {Ziegler}}]{Gressel_etal13}
{Gressel}, O., {Nelson}, R.~P., {Turner}, N.~J., \& {Ziegler}, U. 2013, \apj,
  779, 59

\bibitem[{{Guilet} \& {Ogilvie}(2012)}]{GuiletOgilvie12}
{Guilet}, J. \& {Ogilvie}, G.~I. 2012, \mnras, 424, 2097

\bibitem[{{Guilet} \& {Ogilvie}(2013)}]{GuiletOgilvie13}
---. 2013, \mnras, 430, 822

\bibitem[{{Hartmann} {et~al.}(1998){Hartmann}, {Calvet}, {Gullbring}, \&
  {D'Alessio}}]{Hartmann_etal98}
{Hartmann}, L., {Calvet}, N., {Gullbring}, E., \& {D'Alessio}, P. 1998, \apj,
  495, 385

\bibitem[{{Hayashi}(1981)}]{Hayashi81}
{Hayashi}, C. 1981, Progress of Theoretical Physics Supplement, 70, 35

\bibitem[{{Hennebelle} \& {Ciardi}(2009)}]{HennebelleCiardi09}
{Hennebelle}, P. \& {Ciardi}, A. 2009, \aap, 506, L29

\bibitem[{{Hirose} \& {Turner}(2011)}]{HiroseTurner11}
{Hirose}, S. \& {Turner}, N.~J. 2011, \apjl, 732, L30

\bibitem[{{Hughes} \& {Armitage}(2012)}]{HughesArmitage12}
{Hughes}, A.~L.~H. \& {Armitage}, P.~J. 2012, \mnras, 423, 389

\bibitem[{{Hull} {et~al.}(2014){Hull}, {Plambeck}, {Kwon}, {Bower},
  {Carpenter}, {Crutcher}, {Fiege}, {Franzmann}, {Hakobian}, {Heiles}, {Houde},
  {Hughes}, {Lamb}, {Looney}, {Marrone}, {Matthews}, {Pillai}, {Pound},
  {Rahman}, {Sandell}, {Stephens}, {Tobin}, {Vaillancourt}, {Volgenau}, \&
  {Wright}}]{Hull_etal14}
{Hull}, C.~L.~H., {Plambeck}, R.~L., {Kwon}, W., {Bower}, G.~C., {Carpenter},
  J.~M., {Crutcher}, R.~M., {Fiege}, J.~D., {Franzmann}, E., {Hakobian}, N.~S.,
  {Heiles}, C., {Houde}, M., {Hughes}, A.~M., {Lamb}, J.~W., {Looney}, L.~W.,
  {Marrone}, D.~P., {Matthews}, B.~C., {Pillai}, T., {Pound}, M.~W., {Rahman},
  N., {Sandell}, G., {Stephens}, I.~W., {Tobin}, J.~J., {Vaillancourt}, J.~E.,
  {Volgenau}, N.~H., \& {Wright}, M.~C.~H. 2014, ApJ, submitted

\bibitem[{{Ida} {et~al.}(2008){Ida}, {Guillot}, \& {Morbidelli}}]{Ida_etal08}
{Ida}, S., {Guillot}, T., \& {Morbidelli}, A. 2008, \apj, 686, 1292

\bibitem[{{Igea} \& {Glassgold}(1999)}]{IG99}
{Igea}, J. \& {Glassgold}, A.~E. 1999, \apj, 518, 848

\bibitem[{{Ilgner} \& {Nelson}(2006)}]{IlgnerNelson06}
{Ilgner}, M. \& {Nelson}, R.~P. 2006, \aap, 445, 205

\bibitem[{{Jin}(1996)}]{Jin96}
{Jin}, L. 1996, \apj, 457, 798

\bibitem[{{Johansen} {et~al.}(2009){Johansen}, {Youdin}, \& {Mac
  Low}}]{Johansen_etal09}
{Johansen}, A., {Youdin}, A., \& {Mac Low}, M. 2009, \apjl, 704, L75

\bibitem[{{Joos} {et~al.}(2012){Joos}, {Hennebelle}, \& {Ciardi}}]{Joos_etal12}
{Joos}, M., {Hennebelle}, P., \& {Ciardi}, A. 2012, \aap, 543, A128

\bibitem[{{Klahr} \& {Hubbard}(2014)}]{KlahrHubbard14}
{Klahr}, H. \& {Hubbard}, A. 2014, \apj, 788, 21

\bibitem[{{Kley} \& {Nelson}(2012)}]{KleyNelson12}
{Kley}, W. \& {Nelson}, R.~P. 2012, \araa, 50

\bibitem[{{K{\"o}nigl} {et~al.}(2010){K{\"o}nigl}, {Salmeron}, \&
  {Wardle}}]{Konigl_etal10}
{K{\"o}nigl}, A., {Salmeron}, R., \& {Wardle}, M. 2010, \mnras, 401, 479

\bibitem[{{Krasnopolsky} {et~al.}(2011){Krasnopolsky}, {Li}, \&
  {Shang}}]{Krasnopolsky_etal11}
{Krasnopolsky}, R., {Li}, Z.-Y., \& {Shang}, H. 2011, \apj, 733, 54

\bibitem[{{Kretke} \& {Lin}(2012)}]{KretkeLin12}
{Kretke}, K.~A. \& {Lin}, D.~N.~C. 2012, \apj, 755, 74

\bibitem[{{Kretke} {et~al.}(2009){Kretke}, {Lin}, {Garaud}, \&
  {Turner}}]{Kretke_etal09}
{Kretke}, K.~A., {Lin}, D.~N.~C., {Garaud}, P., \& {Turner}, N.~J. 2009, \apj,
  690, 407

\bibitem[{{Kunz}(2008)}]{Kunz08}
{Kunz}, M.~W. 2008, \mnras, 385, 1494

\bibitem[{{Kunz} \& {Balbus}(2004)}]{KunzBalbus04}
{Kunz}, M.~W. \& {Balbus}, S.~A. 2004, \mnras, 348, 355

\bibitem[{{Kunz} \& {Lesur}(2013)}]{KunzLesur13}
{Kunz}, M.~W. \& {Lesur}, G. 2013, \mnras, 434, 2295

\bibitem[{{Kunz} {et~al.}(2014){Kunz}, {Stone}, \& {Bai}}]{Kunz_etal14}
{Kunz}, M.~W., {Stone}, J.~M., \& {Bai}, X.-N. 2014, Journal of Computational
  Physics, arXiv:1311.4865

\bibitem[{{Latter} \& {Balbus}(2012)}]{LatterBalbus12}
{Latter}, H.~N. \& {Balbus}, S. 2012, \mnras, 424, 1977

\bibitem[{{Latter} {et~al.}(2010){Latter}, {Fromang}, \&
  {Gressel}}]{Latter_etal10}
{Latter}, H.~N., {Fromang}, S., \& {Gressel}, O. 2010, \mnras, 406, 848

\bibitem[{{Lesur} {et~al.}(2013){Lesur}, {Ferreira}, \&
  {Ogilvie}}]{Lesur_etal13}
{Lesur}, G., {Ferreira}, J., \& {Ogilvie}, G.~I. 2013, \aap, 550, A61

\bibitem[{{Lesur} {et~al.}(2014){Lesur}, {Kunz}, \& {Fromang}}]{Lesur_etal14}
{Lesur}, G., {Kunz}, M.~W., \& {Fromang}, S. 2014, ArXiv e-prints

\bibitem[{{Lesur} \& {Papaloizou}(2010)}]{LesurPapaloizou10}
{Lesur}, G. \& {Papaloizou}, J.~C.~B. 2010, \aap, 513, A60

\bibitem[{{Li}(1996)}]{Li96}
{Li}, Z.-Y. 1996, \apj, 465, 855

\bibitem[{{Li} {et~al.}(2011){Li}, {Krasnopolsky}, \& {Shang}}]{Li_etal11}
{Li}, Z.-Y., {Krasnopolsky}, R., \& {Shang}, H. 2011, \apj, 738, 180

\bibitem[{{Lovelace} {et~al.}(1999){Lovelace}, {Li}, {Colgate}, \&
  {Nelson}}]{Lovelace_etal99}
{Lovelace}, R.~V.~E., {Li}, H., {Colgate}, S.~A., \& {Nelson}, A.~F. 1999,
  \apj, 513, 805

\bibitem[{{Lubow} {et~al.}(1994){Lubow}, {Papaloizou}, \&
  {Pringle}}]{Lubow_etal94}
{Lubow}, S.~H., {Papaloizou}, J.~C.~B., \& {Pringle}, J.~E. 1994, \mnras, 268,
  1010

\bibitem[{{Lyra}(2014)}]{Lyra14}
{Lyra}, W. 2014, ArXiv e-prints

\bibitem[{{Lyra} \& {Klahr}(2011)}]{LyraKlahr11}
{Lyra}, W. \& {Klahr}, H. 2011, \aap, 527, A138

\bibitem[{{Marcus} {et~al.}(2013){Marcus}, {Pei}, {Jiang}, \&
  {Hassanzadeh}}]{Marcus_etal13}
{Marcus}, P.~S., {Pei}, S., {Jiang}, C.-H., \& {Hassanzadeh}, P. 2013, Physical
  Review Letters, 111, 084501

\bibitem[{{McElroy} {et~al.}(2013){McElroy}, {Walsh}, {Markwick}, {Cordiner},
  {Smith}, \& {Millar}}]{UMIST12}
{McElroy}, D., {Walsh}, C., {Markwick}, A.~J., {Cordiner}, M.~A., {Smith}, K.,
  \& {Millar}, T.~J. 2013, \aap, arXiv:1212.6362

\bibitem[{{Mellon} \& {Li}(2008)}]{MellonLi08}
{Mellon}, R.~R. \& {Li}, Z.-Y. 2008, \apj, 681, 1356

\bibitem[{{Miyoshi} \& {Kusano}(2005)}]{MiyoshiKusano05}
{Miyoshi}, T. \& {Kusano}, K. 2005, Journal of Computational Physics, 208, 315

\bibitem[{{Nelson} \& {Gressel}(2010)}]{NelsonGressel10}
{Nelson}, R.~P. \& {Gressel}, O. 2010, \mnras, 409, 639

\bibitem[{{Nelson} {et~al.}(2013){Nelson}, {Gressel}, \&
  {Umurhan}}]{Nelson_etal13}
{Nelson}, R.~P., {Gressel}, O., \& {Umurhan}, O.~M. 2013, \mnras, 435, 2610

\bibitem[{{Okuzumi} \& {Hirose}(2011)}]{OkuzumiHirose11}
{Okuzumi}, S. \& {Hirose}, S. 2011, \apj, 742, 65

\bibitem[{{Okuzumi} {et~al.}(2013){Okuzumi}, {Takeuchi}, \&
  {Muto}}]{Okuzumi_etal13}
{Okuzumi}, S., {Takeuchi}, T., \& {Muto}, T. 2013, ArXiv e-prints

\bibitem[{{Ormel} \& {Okuzumi}(2013)}]{OrmelOkuzumi13}
{Ormel}, C.~W. \& {Okuzumi}, S. 2013, \apj, 771, 44

\bibitem[{{O'Sullivan} \& {Downes}(2006)}]{OSullivanDownes06}
{O'Sullivan}, S. \& {Downes}, T.~P. 2006, \mnras, 366, 1329

\bibitem[{{O'Sullivan} \& {Downes}(2007)}]{OSullivanDownes07}
---. 2007, \mnras, 376, 1648

\bibitem[{{Paardekooper} {et~al.}(2011){Paardekooper}, {Baruteau}, \&
  {Kley}}]{Paardekooper_etal11}
{Paardekooper}, S.-J., {Baruteau}, C., \& {Kley}, W. 2011, \mnras, 410, 293

\bibitem[{{Perez-Becker} \&
  {Chiang}(2011{\natexlab{a}})}]{PerezBeckerChiang11b}
{Perez-Becker}, D. \& {Chiang}, E. 2011{\natexlab{a}}, \apj, 735, 8

\bibitem[{{Perez-Becker} \&
  {Chiang}(2011{\natexlab{b}})}]{PerezBeckerChiang11a}
---. 2011{\natexlab{b}}, \apj, 727, 2

\bibitem[{{Petersen} {et~al.}(2007){Petersen}, {Stewart}, \&
  {Julien}}]{Petersen_etal07b}
{Petersen}, M.~R., {Stewart}, G.~R., \& {Julien}, K. 2007, \apj, 658, 1252

\bibitem[{{Pinilla} {et~al.}(2012){Pinilla}, {Birnstiel}, {Ricci}, {Dullemond},
  {Uribe}, {Testi}, \& {Natta}}]{Pinilla_etal12}
{Pinilla}, P., {Birnstiel}, T., {Ricci}, L., {Dullemond}, C.~P., {Uribe},
  A.~L., {Testi}, L., \& {Natta}, A. 2012, \aap, 538, A114

\bibitem[{{Preibisch} {et~al.}(2005){Preibisch}, {Kim}, {Favata}, {Feigelson},
  {Flaccomio}, {Getman}, {Micela}, {Sciortino}, {Stassun}, {Stelzer}, \&
  {Zinnecker}}]{Preibisch_etal05}
{Preibisch}, T., {Kim}, Y., {Favata}, F., {Feigelson}, E.~D., {Flaccomio}, E.,
  {Getman}, K., {Micela}, G., {Sciortino}, S., {Stassun}, K., {Stelzer}, B., \&
  {Zinnecker}, H. 2005, \apjs, 160, 401

\bibitem[{{Raettig} {et~al.}(2013){Raettig}, {Lyra}, \&
  {Klahr}}]{Raettig_etal13}
{Raettig}, N., {Lyra}, W., \& {Klahr}, H. 2013, \apj, 765, 115

\bibitem[{{Ribas} {et~al.}(2013){Ribas}, {Mer{\'{\i}}n}, {Bouy}, \&
  {Maud}}]{Ribas_etal14}
{Ribas}, {\'A}., {Mer{\'{\i}}n}, B., {Bouy}, H., \& {Maud}, L.~T. 2013, ArXiv
  e-prints

\bibitem[{{Salmeron} {et~al.}(2011){Salmeron}, {K{\"o}nigl}, \&
  {Wardle}}]{Salmeron_etal11}
{Salmeron}, R., {K{\"o}nigl}, A., \& {Wardle}, M. 2011, \mnras, 412, 1162

\bibitem[{{Sano} \& {Miyama}(1999)}]{SanoMiyama99}
{Sano}, T. \& {Miyama}, S.~M. 1999, \apj, 515, 776

\bibitem[{{Sano} {et~al.}(2000){Sano}, {Miyama}, {Umebayashi}, \&
  {Nakano}}]{Sano_etal00}
{Sano}, T., {Miyama}, S.~M., {Umebayashi}, T., \& {Nakano}, T. 2000, \apj, 543,
  486

\bibitem[{{Sano} \& {Stone}(2002{\natexlab{a}})}]{SanoStone02a}
{Sano}, T. \& {Stone}, J.~M. 2002{\natexlab{a}}, \apj, 570, 314

\bibitem[{{Sano} \& {Stone}(2002{\natexlab{b}})}]{SanoStone02b}
---. 2002{\natexlab{b}}, \apj, 577, 534

\bibitem[{{Santos-Lima} {et~al.}(2012){Santos-Lima}, {de Gouveia Dal Pino}, \&
  {Lazarian}}]{SantosLima_etal12}
{Santos-Lima}, R., {de Gouveia Dal Pino}, E.~M., \& {Lazarian}, A. 2012, \apj,
  747, 21

\bibitem[{{Seifried} {et~al.}(2012){Seifried}, {Banerjee}, {Pudritz}, \&
  {Klessen}}]{Seifried_etal12}
{Seifried}, D., {Banerjee}, R., {Pudritz}, R.~E., \& {Klessen}, R.~S. 2012,
  \mnras, 423, L40

\bibitem[{{Shakura} \& {Sunyaev}(1973)}]{ShakuraSunyaev73}
{Shakura}, N.~I. \& {Sunyaev}, R.~A. 1973, \aap, 24, 337

\bibitem[{{Sicilia-Aguilar} {et~al.}(2006){Sicilia-Aguilar}, {Hartmann},
  {Calvet}, {Megeath}, {Muzerolle}, {Allen}, {D'Alessio}, {Mer{\'{\i}}n},
  {Stauffer}, {Young}, \& {Lada}}]{Sicilia_etal06}
{Sicilia-Aguilar}, A., {Hartmann}, L., {Calvet}, N., {Megeath}, S.~T.,
  {Muzerolle}, J., {Allen}, L., {D'Alessio}, P., {Mer{\'{\i}}n}, B.,
  {Stauffer}, J., {Young}, E., \& {Lada}, C. 2006, \apj, 638, 897

\bibitem[{{Simon} {et~al.}(2013{\natexlab{a}}){Simon}, {Bai}, {Armitage},
  {Stone}, \& {Beckwith}}]{Simon_etal13b}
{Simon}, J.~B., {Bai}, X.-N., {Armitage}, P.~J., {Stone}, J.~M., \& {Beckwith},
  K. 2013{\natexlab{a}}, \apj, 775, 73

\bibitem[{{Simon} {et~al.}(2013{\natexlab{b}}){Simon}, {Bai}, {Stone},
  {Armitage}, \& {Beckwith}}]{Simon_etal13a}
{Simon}, J.~B., {Bai}, X.-N., {Stone}, J.~M., {Armitage}, P.~J., \& {Beckwith},
  K. 2013{\natexlab{b}}, \apj, 764, 66

\bibitem[{{Stapelfeldt} {et~al.}(2003){Stapelfeldt}, {M{\'e}nard}, {Watson},
  {Krist}, {Dougados}, {Padgett}, \& {Brandner}}]{Stapelfeldt_etal03}
{Stapelfeldt}, K.~R., {M{\'e}nard}, F., {Watson}, A.~M., {Krist}, J.~E.,
  {Dougados}, C., {Padgett}, D.~L., \& {Brandner}, W. 2003, \apj, 589, 410

\bibitem[{{Stone} \& {Gardiner}(2010)}]{StoneGardiner10}
{Stone}, J.~M. \& {Gardiner}, T.~A. 2010, \apjs, 189, 142

\bibitem[{{Stone} {et~al.}(2008){Stone}, {Gardiner}, {Teuben}, {Hawley}, \&
  {Simon}}]{Stone_etal08}
{Stone}, J.~M., {Gardiner}, T.~A., {Teuben}, P., {Hawley}, J.~F., \& {Simon},
  J.~B. 2008, \apjs, 178, 137

\bibitem[{{Suzuki} \& {Inutsuka}(2009)}]{SuzukiInutsuka09}
{Suzuki}, T.~K. \& {Inutsuka}, S.-i. 2009, \apjl, 691, L49

\bibitem[{{Suzuki} \& {Inutsuka}(2014)}]{SuzukiInutsuka14}
---. 2014, \apj, 784, 121

\bibitem[{{Takeuchi} \& {Okuzumi}(2013)}]{TakeuchiOkuzumi13}
{Takeuchi}, T. \& {Okuzumi}, S. 2013, ArXiv e-prints

\bibitem[{{Tomida} {et~al.}(2013){Tomida}, {Tomisaka}, {Matsumoto}, {Hori},
  {Okuzumi}, {Machida}, \& {Saigo}}]{Tomida_etal13}
{Tomida}, K., {Tomisaka}, K., {Matsumoto}, T., {Hori}, Y., {Okuzumi}, S.,
  {Machida}, M.~N., \& {Saigo}, K. 2013, \apj, 763, 6

\bibitem[{{T{\'o}th} {et~al.}(2008){T{\'o}th}, {Ma}, \&
  {Gombosi}}]{Toth_etal08}
{T{\'o}th}, G., {Ma}, Y., \& {Gombosi}, T.~I. 2008, Journal of Computational
  Physics, 227, 6967

\bibitem[{{Turner} {et~al.}(2014){Turner}, {Fromang}, {Gammie}, {Lesur},
  {Wardle}, \& {Bai}}]{Turner_etal14}
{Turner}, N.~J., {Fromang}, S., {Gammie}, C.~F., {Lesur}, G., {Wardle}, M., \&
  {Bai}, X.-N. 2014, in PPVI, ed. C.~P. {Dullemond} No. arXiv:1401.7306

\bibitem[{{Turner} \& {Sano}(2008)}]{TurnerSano08}
{Turner}, N.~J. \& {Sano}, T. 2008, \apjl, 679, L131

\bibitem[{{Urpin}(2003)}]{Urpin03}
{Urpin}, V. 2003, \aap, 404, 397

\bibitem[{{Varni{\`e}re} \& {Tagger}(2006)}]{VarniereTagger06}
{Varni{\`e}re}, P. \& {Tagger}, M. 2006, \aap, 446, L13

\bibitem[{{Walsh} {et~al.}(2010){Walsh}, {Millar}, \& {Nomura}}]{Walsh_etal10}
{Walsh}, C., {Millar}, T.~J., \& {Nomura}, H. 2010, \apj, 722, 1607

\bibitem[{{Walsh} {et~al.}(2012){Walsh}, {Nomura}, {Millar}, \&
  {Aikawa}}]{Walsh_etal12}
{Walsh}, C., {Nomura}, H., {Millar}, T.~J., \& {Aikawa}, Y. 2012, \apj, 747,
  114

\bibitem[{{Wardle}(1997)}]{Wardle97}
{Wardle}, M. 1997, in Astronomical Society of the Pacific Conference Series,
  Vol. 121, IAU Colloq. 163: Accretion Phenomena and Related Outflows, ed.
  D.~T. {Wickramasinghe}, G.~V. {Bicknell}, \& L.~{Ferrario}, 561

\bibitem[{{Wardle}(1999)}]{Wardle99}
{Wardle}, M. 1999, \mnras, 307, 849

\bibitem[{{Wardle}(2007)}]{Wardle07}
---. 2007, \apss, 311, 35

\bibitem[{{Wardle} \& {Koenigl}(1993)}]{WardleKoenigl93}
{Wardle}, M. \& {Koenigl}, A. 1993, \apj, 410, 218

\bibitem[{{Wardle} \& {Salmeron}(2012)}]{WardleSalmeron12}
{Wardle}, M. \& {Salmeron}, R. 2012, \mnras, 422, 2737

\bibitem[{{Weidenschilling}(1977)}]{Weidenschilling77}
{Weidenschilling}, S.~J. 1977, \mnras, 180, 57

\bibitem[{{Wolk} {et~al.}(2005){Wolk}, {Harnden}, {Flaccomio}, {Micela},
  {Favata}, {Shang}, \& {Feigelson}}]{Wolk_etal05}
{Wolk}, S.~J., {Harnden}, Jr., F.~R., {Flaccomio}, E., {Micela}, G., {Favata},
  F., {Shang}, H., \& {Feigelson}, E.~D. 2005, \apjs, 160, 423

\bibitem[{{Yang} {et~al.}(2012){Yang}, {Mac Low}, \& {Menou}}]{Yang_etal12}
{Yang}, C.-C., {Mac Low}, M.-M., \& {Menou}, K. 2012, \apj, 748, 79

\bibitem[{{Youdin}(2011)}]{Youdin11}
{Youdin}, A.~N. 2011, \apj, 731, 99

\end{thebibliography}

\label{lastpage}
\end{document}